\documentclass[acmsmall]{acmart}
\AtEndDocument{\clearpage\setcounter{TotPages}{27}}

\usepackage{setup}
\author{Mickaël Laurent}
\email{mickael.laurent@matfyz.cuni.cz}
\orcid{0000-0003-1590-2392}
\affiliation{
  \institution{Charles University}
  \city{Prague}
  \country{Czech Republic}
} 
\authornote{Both authors contributed equally to the paper}

\author{Kim Nguy\~{ê}n}
\email{kn@lmf.cnrs.fr}
\orcid{0000-0002-1729-870X}
\affiliation{
  \institution{Université Paris-Saclay, CNRS, ENS Paris-Saclay, Laboratoire Méthodes Formelles}
  \city{Gif-sur-Yvette}
  \country{France}
}
\authornotemark[1]

\setcopyright{cc}
\setcctype{by}
\acmDOI{10.1145/3839466}
\acmYear{2026}
\acmJournal{PACMPL}
\acmVolume{10}
\acmNumber{OOPSLA2}
\acmArticle{334}
\acmMonth{10}



\begin{document}

\begin{tikzpicture}[remember picture, overlay]
\node at ($(current page.north east) + (-4cm,-2cm)$) {%
\includegraphics[width=2cm]{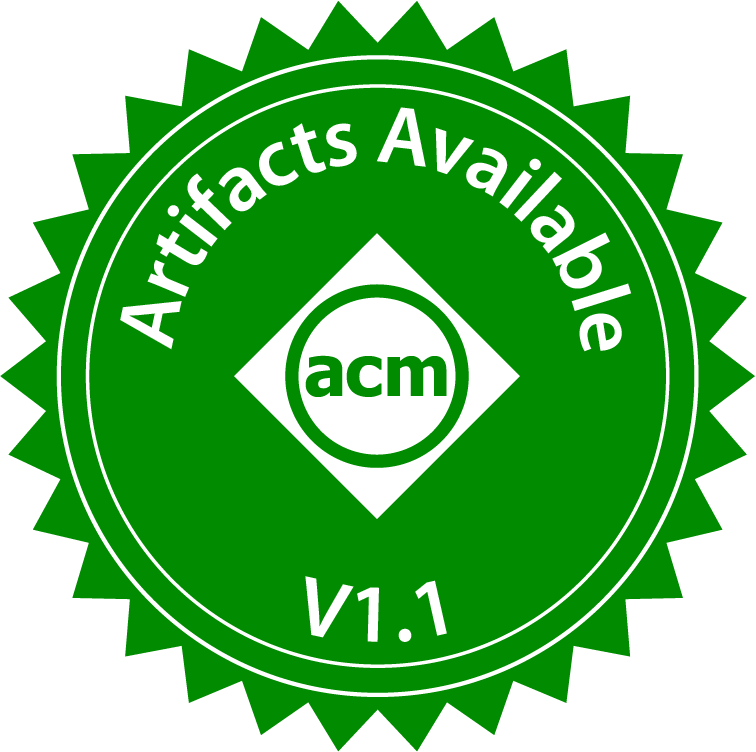}%
\includegraphics[width=2cm]{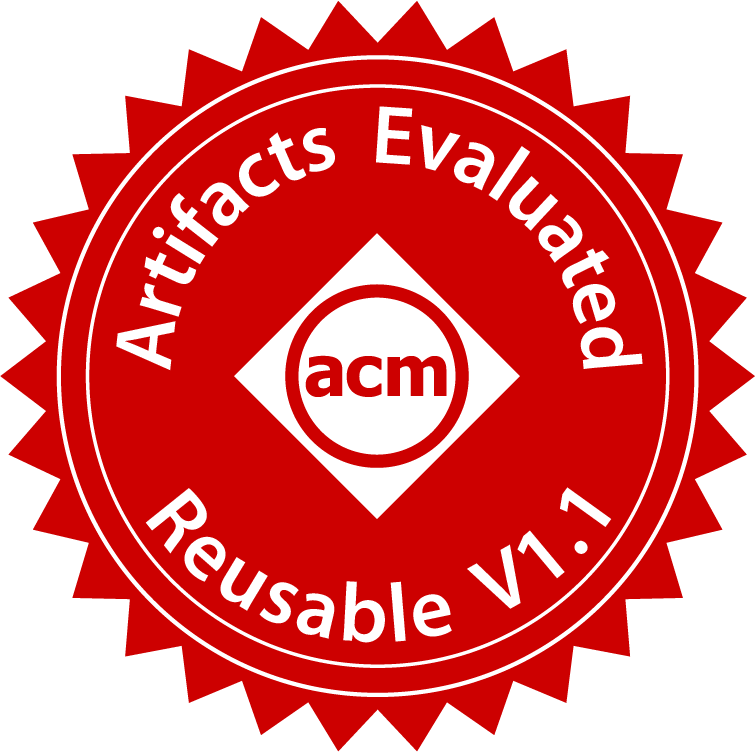}%
\includegraphics[width=2cm]{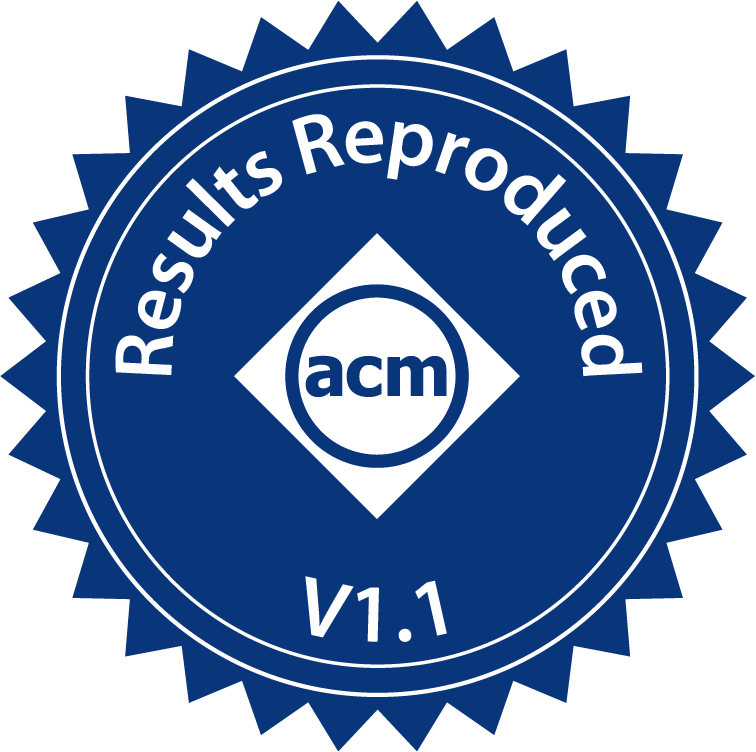}%
};
\end{tikzpicture}
\title{Implementing Set-Theoretic Types}

\begin{abstract}
Set-theoretic types provide a rich type algebra that supports unrestricted unions, intersections, and negations,
together with a decidable type constraint-solving algorithm known as tallying. These types are particularly well
suited for typing dynamic languages, where functions often exhibit both generic and overloaded behavior.
However, the complexity of their implementation has hindered their widespread adoption.
In this paper, we introduce a modular representation for set-theoretic types and revisit the algorithms for subtyping and tallying.
We compare our approach with the historical CDuce implementation and evaluate the performance impact of some optimizations
and design choices.
\end{abstract}
\maketitle

\section{Introduction}\label{sec:intro}
For over a decade, programmers have added static type systems to dynamic, originally
untyped languages such as JavaScript and Python. These efforts aim to combine
the flexibility of dynamic typing with the reliability of static type checking.
Examples of such type systems are \href{https://www.typescriptlang.org/}{TypeScript} and
\href{https://flow.org/}{Flow} for JavaScript or \href{https://mypy.readthedocs.io/en/stable/}{Mypy}, \href{https://pyre-check.org/}{Pyre} and \href{https://microsoft.github.io/pyright/}{Pyright} for Python. To
account for the variety of programming patterns expressible in untyped
languages, these type systems often feature several (if not all) of the typing constructs
that are part of the programming language literature, namely:
\begin{itemize}
  \item parametric polymorphism (often referred to as \emph{generics})
  \item overloading or ad-hoc polymorphism, together with subtyping
  \item occurrence typing~\cite{tobin2008} 
  (the refinement of a type following a dynamic type test)
  \item arbitrary union types (e.g. to express heterogeneous collections or optional results)
  \item intersection types (in particular for functions or records)
  \item gradual typing~\cite{gradual}, the ability to mix typed and untyped code
\end{itemize}
Mixing together \emph{all} of these, in a principled way, is a
daunting task.  Most, if not all, of the previously mentioned systems
do not guarantee type safety and mix the various typing constructs in
an ad-hoc way. Yet, for some of their use cases (documentation, test
generation, code completion in IDEs, …) they provide excellent
performance, and are able to infer \emph{some} type information
relatively quickly.

A theoretical formalism handling these features does exist: \emph{set-theoretic
types}. They were first introduced in the context of XML programming
\cite{xduce,AlainThesis} and later extended with parametric
polymorphism~\cite{polyduce2}, gradual typing~\cite{CLPS19} and type narrowing
\cite{popl24}. While set-theoretic types have started to be used as a
theoretical foundation for type systems (e.g.
\href{https://github.com/etylizer/etylizer}{Etylizer}~\cite{erlang} for Erlang,
\href{https://github.com/elixir-lang/elixir/tree/main/lib/elixir/lib/module/types}{Elixir}~\cite{elixir}),
they have not seen the wide adoption one could have hoped for. We believe that
one of the main issues is the disconnect between the mathematically elegant
theoretical foundation and the design and implementation issues a practical
implementation faces.

The goal of this work is to report on SSTT,
the simple set-theoretic type library, a reference implementation of
set-theoretic types. We cover the implementation and data-structure of
the basic type algebra, show how to extend it to cover
practical (but often ignored) data types, and present
implementation techniques of high-level operators, namely
subtyping and tallying (computing a set of substitutions that solve a
subtyping constraint).
%

\subsection{Overview}

One of the main features of dynamic languages is their ability to manipulate data
of heterogeneous types. A key construct that makes it possible is the \textit{typecase},
a dynamic test of type that can be used to encode overloaded functions (i.e. functions
whose behavior depends on the type of their arguments).
For instance, the function $\lambda x.\;\tcase{x}{\Int}{x+1}{x}$ tests whether its argument
$x$ is an integer, in which case it returns $x+1$. Otherwise, it returns the argument $x$ unmodified.
To type such a function, the set connectives provided by set-theoretic types are of great use.
Assuming the parameter $x$ has type $\Any$ (the top type, inhabited by every value of the language),
a type-checker may be able to deduce that, when typing the first branch $x+1$ of the type-case,
the type of $x$ can be safely intersected with $\Int$. Likewise, when typing the second branch
$x$, the type of $x$ can be safely intersected with $\neg\Int$. This is usually referred to as
\emph{type narrowing} or \emph{occurrence typing}~\cite{THF10, popl24}. Assuming $+$ is of type
$\Int\to\Int\to\Int$, we get the type $\Int$ for the first branch, and $\neg\Int$ for the second
branch: the resulting type for the function is thus $\Int \vee \neg\Int$,
which in the set-theoretic type theory is equivalent to $\Any$.

Though $\Any\to\Any$ is a correct type for this function,
a type-checker may be able to infer more precise types. For instance, by choosing the domain
$\Int$ for $x$, it could infer the type $\Int\to\Int$ (the second branch of the typecase is unreachable).
Likewise, by choosing the domain $\neg\Int$, it could infer the type $\neg\Int \to \neg\Int$.
Set-theoretic types allow us to capture both behaviors using an intersection type:
our function can be given the type $(\Int\to\Int)\land(\neg\Int \to \neg\Int)$.

Going further, we may choose to also capture the genericity of the function:
when given an argument that is not an integer, this argument is returned, unchanged.
To do so, we can type our function for the domain $\alpha\setminus\Int$
(shorthand for $\alpha\land\neg\Int$), inferring the type $\alpha\setminus\Int\to\alpha\setminus\Int$
where $\alpha$ is a \emph{type variable} that can freely be substituted by any type.
Combining this with our earlier type, we can infer for this function the type
$(\Int\to\Int)\land(\alpha\setminus\Int\to\alpha\setminus\Int)$, which combines both
\emph{parametric polymorphism} (genericity) and \emph{ad-hoc polymorphism} (overloading).

As we have just seen, generic and overloaded functions can be typed using arrow types,
type variables, and intersections. Let us now focus on the typing of applications involving such functions.
We use the following example from \cite{polyduce2}.

\noindent
\begin{minipage}[t]{.98\textwidth}
\begin{stmlcode}
  val map : $\tscheme{\alpha,\beta}{(\alpha\to\beta)\to\ListS{\alpha}}\to\ListS{\beta}$
  val even : $\tscheme{\gamma}{(\Int\to\Bool)\wedge((\gamma\setminus\Int) \to (\gamma\setminus\Int))}$
\end{stmlcode}
  \end{minipage}

\noindent Here, the \code{map} function has its familiar type, with the twist that list
types are \emph{regular expression types}. These are merely syntactic sugar for
recursive types. For instance, the type $\ListS{\alpha}$ is the
recursive type defined by the equation $X = {\Nil}~\vee~\pair{\alpha}{X}$, which
is a \emph{union} of $\Nil$ --- the \emph{singleton} type of the constant
representing the empty list --- and a product type of the type variable
$\alpha$ (the polymorphic type of the head of the list) and $X$ standing for the
list type itself (the tail of the list). The \code{even} function is an
\emph{intersection} of two arrow types, meaning that \code{even} is an
overloaded function. When this function is applied to a value of type \Int, it
returns a value of type \Bool. Otherwise, when applied to values that are not
integers, captured by the \emph{set difference} $\gamma\setminus\Int$, it
returns a value of that type\footnote{To implement such behavior, the
underlying programming language supports a dynamic type case construct.}.
Now, let's imagine what would be needed to
type the (partial) application \code{map even}. In a
Hindley-Milner style type system, we would need to:
\begin{enumerate}
  \item find a type substitution which unifies the type of the domain of
  \code{map} with the type of \code{even}
  \item apply this substitution to 
  ($\ListS{\alpha}\to\ListS{\beta}$) to deduce the type
  of the whole expression.
\end{enumerate}
However, in the context of set-theoretic types, \emph{syntactic} unification is
not powerful enough. Indeed, it would fail here, since it
confronts an arrow type ($\alpha\to\beta$) against an intersection type (the
type of \code{even}). One of the key features of set-theoretic types is the
associated notion of \emph{semantic subtyping}. With this notion, the behavior
of set-theoretic connectives w.r.t. type constructors (e.g. idempotence, distributivity)
is derived from set theory. To
reconcile parametric polymorphism and subtyping-based polymorphism,
\citet{polyduce2} introduce the \emph{tallying} operation. In a nutshell,
tallying consists in finding type substitutions for type variables that make a
set of type inequations hold. Thus, typing the application \code{map even}
consists in finding type substitutions such that
\begin{displaymath}
  (\alpha\to\beta) \to\ListS{\alpha}\to\ListS{\beta} \leq
  ((\Int\to\Bool)\wedge((\gamma\setminus\Int) \to (\gamma\setminus\Int)))\to\delta
\end{displaymath}
where $\leq$ denotes subtyping and $\delta$ is a freshly introduced type
variable that represents the type we are interested in, that is, the type of
\code{map even}. This single problem contains the constraints that:
\begin{itemize}
  \item the type of \code{even} must be a subtype of $\alpha\to\beta$ (contravariance on the domain of \code{map})
  \item $\delta$ must be a supertype of
  $\ListS{\alpha}\to\ListS{\beta}$ (covariance on the codomain of \code{map})
\end{itemize}
The result of this tallying problem is a set of four substitutions $\{\sigma_1,\sigma_2,\sigma_3,\sigma_4\}$,
and the type of the application is therefore $\delta\sigma_1\wedge\delta\sigma_2\wedge\delta\sigma_3\wedge\delta\sigma_4$,
which  becomes the type of the overloaded function:
\begin{displaymath}
  \begin{array}{c}
    (\ListS{(\Int\vee\gamma_1)}\to\ListS{(\Bool\vee(\gamma_1\setminus\Int))}) ~~\wedge~~
    (\nil\to\nil) ~~\wedge\\
(\ListS{(\gamma_3\setminus\Int)}\to\ListS{(\gamma_3\setminus\Int)})~~\wedge~~
(\ListS{\Int}\to\ListS{\Bool})
  \end{array}
\end{displaymath}
This short example highlights most of the challenges one faces when trying to
implement set-theoretic types. While some of them are documented in the
literature, we propose in this work a systematic presentation of
all the relevant aspects of the implementation of set-theoretic types, together
with novel implementation techniques and insights discovered while implementing
the SSTT library. Each technique is evaluated on a realistic use of
set-theoretic types.

\subsection{Contributions} 
The contributions of this paper are as follows:
\begin{itemize}
  \item  We introduce \emph{Binary Decision Trees} (BDTs),
  a data-structure used to efficiently represent Boolean formulas involving custom atoms.
  We define a novel semantic simplification operation that reduces a BDT based on a semantic
  equivalence relation over the atoms (Section~\ref{sec:bdd}).
  \item We present a modular design for implementing set-theoretic types. We
  choose to represent them using a layered API, each layer handling a specific
  feature of set-theoretic types: type constructors, set-theoretic
  operators, type variables, recursivity (Section~\ref{sec:nodes}).

  \item We revisit the subtyping and tallying algorithms
  and propose new optimizations (Section~\ref{sec:subtyping-tallying}).
  \item We present several useful extensions that can be encoded in the base type algebra (Section~\ref{sec:extensions}).
  \item We present our SSTT library,
  compare it against the CDuce implementation, and
  evaluate the impact of the semantic simplification and various optimizations (Section~\ref{sec:eval}).
\end{itemize}
We recall the basics of set-theoretic types in Section~\ref{sec:defs}
and review the related work in Section~\ref{sec:related}.


\section{Set-Theoretic Types}
\label{sec:defs}
\subsection{Definitions}
The types we implement in this paper are the set-theoretic types. For more
detail, we refer the reader to the survey paper \cite{Cas22} which synthesizes
the relevant results from the literature.

\begin{definition}[Set-theoretic types]\label{def:types}
  The set $\Types$ of \emph{set-theoretic types} is the set of
  regular and contractive terms coinductively defined by the following grammar:
\begin{equation*}
  \begin{array}{lrclr}
    \textbf{Types}  &t &::=& b\alt \tv\alt t\to t\alt \pair t t
    \alt t\vee t\alt t\wedge t \alt \neg t \alt \Empty\alt \Any
  \end{array}
\end{equation*}
where $b\in\BaseTypes$ is a base type (in particular, it includes
the constants $c\in\Constants$ of the language we want to type) and $\tv\in\Vars$ is a type variable.
The notation $t_1\setminus t_2$ is a syntactic sugar for $t_1\land\neg t_2$.
When writing a term,
we use the following precedence (by decreasing priority): $\neg$, $\setminus$, $\wedge$, $\vee$, $\pair{}{}$, $\to$.
\end{definition}

As they are defined coinductively, types can be infinite trees, provided that they satisfy the
constraints of regularity and contractivity explained below to ensure decidability of the subtyping relation.
This yields a definition of equirecursive types that does not require explicit binders for recursion.
A term is said \emph{regular} if it only has a finite number of distinct subterms,
and \emph{contractive} if every infinite branch goes through an infinite number of
arrows and products ($\to$ and $\pair{}{}$).

The type $\Empty$ is a special type that is not inhabited by any value,
and is the subtype of all types. Conversely, the type $\Any$ is the supertype of all types.
The $\pair{}{}$ constructor is used to type pairs, and the $\to$ constructor is used to type functions.
These are the most common type constructors, but set-theoretic types can be extended with
other constructors as we will see throughout this paper.

Set-theoretic types are equipped with a decidable subtyping relation $\leq$
(referred to as \emph{semantic subtyping}, cf. Appendix~\ref{app:interpretation}
for more details). For this presentation, it suffices to consider that each type
can be interpreted as a set of values that have that type, and that subtyping is
set containment. Type connectives (union, intersection, negation) are
interpreted as the corresponding set-theoretic operators. Although the interpretation
of type variables is more complex, the following property is sufficient to grasp the behavior
of subtyping in presence of type variables:
\begin{proposition}[Subtyping, \cite{polysub}]\label{prop:subst} Let $t$ and $s$ be two types. The type $s$ is a subtype of type $t$
  if, for every type substitution $\sigma$, we have $s\sigma \leq t\sigma$.
\end{proposition}
We note $\simeq$ the semantic equivalence:
$t_1\simeq t_2$ if and only if $t_1\leq t_2 \text{ and } t_2 \leq t_1$.
Thanks to negation, the subtyping problem is equivalent to checking the emptiness of a type:
\(
  ~~~ s \leq t  ~~~\Longleftrightarrow~~~s\setminus t\leq\Empty
\).

As presented in the introduction, subtyping is a building block for the definition of tallying:
\begin{definition}[Tallying, \cite{polyduce2}]
  Let $S=\{ (s_1, t_1), \ldots, (s_n, t_n)\}$ be a finite set of pairs of types
  and $\Delta$ a set of type variables. The solution of the tallying problem for $S$ and $\Delta$ is:
  \begin{displaymath}
    \ftally(S,\Delta) = \{ \sigma ~|~ \dom{\sigma}  \cap\Delta=\varnothing ~\land~ \forall (s,t)\in S.\; s\sigma\leq t\sigma \}
  \end{displaymath}
\end{definition}
The set $S$ contains the subtyping constraints, and the set $\Delta$ is the set of
variables that the tallying is not allowed to instantiate. An important
property of the tallying operation is that it is complete: it computes a finite
set of substitutions which exactly characterize all the possible solutions.

\subsection{Types in Disjunctive Normal Forms}\label{sec:dnf}
As hinted to in the previous section, deciding subtyping is equivalent to
checking the emptiness of a type. A convenient way to express the subtyping
algorithm is to put the type in \emph{disjunctive normal form} (DNF).
Given a type $t$, its DNF as defined in \cite{polysub} is a syntactic term of the following form:
\begin{displaymath}
  \textsf{DNF}(t) =
  \begin{array}{clr}
    &\bigvee~~\bigwedge_i \alpha^\textsf{b}_i~\wedge~\bigwedge_j{\lnot\beta^\textsf{b}_j}~~\wedge~~\bigwedge_k b_k~\wedge~\bigwedge_l{\lnot b_l}\\
\vee &\bigvee~~\bigwedge_i \alpha^\textsf{p}_i~\wedge~\bigwedge_j{\lnot\beta^\textsf{p}_j}~~\wedge~~\bigwedge_k (t^1_k\times t^2_k)~\wedge~\bigwedge_l{\lnot (s^1_l\times s^2_l)}&\label{eqdnf}\tag{*}\\
\vee &\bigvee~~\bigwedge_i \alpha^\textsf{a}_i~\wedge~\bigwedge_j{\lnot\beta^\textsf{a}_j}~~\wedge~~\bigwedge_k (t^1_k\to t^2_k)~\wedge~\bigwedge_l{\lnot (s^1_l\to s^2_l)}
  \end{array}
\end{displaymath}
In this formula, each row is itself a DNF of positive and negative type
variables and positive and negative instances of a particular type constructor:
a basic type (we assume for now a single kind of basic type), a product, or an
arrow. The subtyping algorithm can then iterate through the elements of this DNF
to test the emptiness of the type. It is known that the complexity of semantic
subtyping is in $2^{O(n)}$ where $n$ is the size of the type (see
\cite{Gesbert2015}). We attempt to tame this complexity in practice both at the
level of type representation and in the subtyping and tallying algorithm.

\section{Binary Decision Trees}\label{sec:bdd}

In this section, we describe \emph{Binary Decision Trees} (BDTs), a data
structure for representing Boolean combinations of \emph{atoms}. The nature of
these atoms is varied, as shown in the previous section. They can be type
variables, basic types, or type constructors. We develop the meta-theory of BDT
and operations by abstracting over atoms. In addition to the standard Boolean
operations on BDTs (Section~\ref{sec:bdd-st}), we define a semantic
simplification operation (Section~\ref{sec:bdd-simpl}).

\subsection{Structure of a BDT}\label{sec:bdd-struct}

Let $\Atoms$ be a set of atoms, ranged over by $\atom$.
We assume we have a total order $\tleq$ over atoms,
as well as an interpretation $\Bint{\atom}$ of an atom $\atom$ as a type.
This total order $\tleq$ can be any arbitrary order, even though it influences the
structure (and size) of BDTs: in practice, it is not a significant factor in
the complexity of set-theoretic operations in comparison to the subtyping and tallying algorithms,
as we observe in Section~\ref{sec:eval}.
Let $\Leaves$ be a set of leaves, ranged over by $\leaf$.
We assume we have an interpretation $\Bint \leaf$ of a leaf $\leaf$ as a type,
a bottom leaf element $\LE$ whose interpretation $\Bint \LE$ is the type $\Empty$,
and a top leaf element $\LA$ whose interpretation $\Bint \LA$ is the type $\Any$.
In addition, the set-theoretic operations $\land$, $\lor$, and $\neg$ must be defined on leaves,
in accordance with the interpretation $\Bint{.}$
($\Bint{\leaf_1 \land \leaf_2}\simeq \Bint{\leaf_1}\land \Bint{\leaf_2}$,\;\;\;
$\Bint{\leaf_1 \lor \leaf_2}\simeq \Bint{\leaf_1}\lor \Bint{\leaf_2}$,\;\;\;
$\Bint{\neg\leaf}\simeq\neg\Bint{\leaf}$).

\begin{definition}[BDT]
  The set of Binary Decision Trees $\BDTs{\Atoms}{\Leaves}$
  over atoms $\Atoms$ and leaves $\Leaves$ are the finite trees produced by the following grammar
  (where $\leaf\in\Leaves$ and $\atom\in\Atoms$):
\begin{displaymath}
  \begin{array}{lrrl}
    \textbf{(Ordered) Binary Decision Trees} &\B &::=& \BL{\leaf} \alt \BN{\atom}{\B}{\B}
  \end{array}
\end{displaymath}
and satisfying the following properties:\\
\textit{(ordering)} for any non-root node labeled $\atom$ whose parent is labeled $\atom'$, we have $\atom\not\tleq\atom'$, and\\
\textit{(reduction)} all non leaf subtrees of the BDT are of the form
$\BN{\atom}{\Bp}{\Bm}$ with $\Bp\neq\Bm$. 
\end{definition}

In particular, the \textit{(ordering)} property guarantees no branch goes through the same label twice; the \textit{(reduction)} property ensures a BDT does not represent a trivially empty type (cf. Definition~\ref{def:bdt-inter}).
Note that it is always possible to create BDTs that are reduced. To
do so, we assume that  $\BN{\_}{\_}{\_}$ behaves as a \emph{smart constructor} which
performs the following simplification: $\BN{\atom}{B}{B} \leadsto B$.

For concision, we use the notation $\BE$ for $\BL{\LE}$, $\BA$ for $\BL{\LA}$,
and $\BS{\atom}$ for the BDT node $\BN{\atom}{\BA}{\BE}$.
The auxiliary definitions below allow us
to extract the set of atoms or the set of leaves of a BDT:
\begin{definition}
  For any BDT $\B$, we define $\Bleaves{\B}$ and $\Batoms{\B}$ as follows:
\[\begin{array}{rclcrcl}
  \Bleaves{\BL{\leaf}} &=& \{\leaf\} &\hspace*{5mm}&
  \Bleaves{\BN{\atom}{\Bp}{\Bm}} &=& \Bleaves{\Bp} \cup \Bleaves{\Bm}\\
  \Batoms{\BL\leaf} &=& \varnothing &&
  \Batoms{\BN{\atom}{\Bp}{\Bm}} &=& \Batoms{\Bp} \cup \Batoms{\Bm} \cup \{\atom\}\\
\end{array}\]
\end{definition}
The meaning of a BDT, that is, the type that it represents, is given by an interpretation function
from BDTs to types.
\begin{definition}\label{def:bdt-inter}
  The \emph{interpretation} $\Bint \B$ of a binary decision tree $\B$ is inductively defined as follows:
\[\begin{array}{rclcrcl}
  \Bint {\BL \leaf} &=& \Bint \leaf &\hspace*{5mm}&
  \Bint {\BN{\atom}{\Bp}{\Bm}} &=& (\Bint\atom\land \Bint \Bp)\vee (\neg \Bint\atom \land \Bint \Bm)
\end{array}\]
\end{definition}
Note that the interpretation of a BDT $\B$ is a Boolean combination of atoms and leaves.
An example of a BDT and the associated interpretation can be found in Figure~\ref{fig:bdd}.
This Boolean combination can be expressed as a DNF (cf. Section~\ref{sec:dnf}):
\begin{definition} The DNF $\Bdnf{\B}$ of a BDT $\B$ is a set of triples $(A_p, A_n, \leaf)$,
where each triple represents a clause intersecting the positive atoms $A_p$, the negative atoms $A_n$, and the leaf $\leaf$:
\[\begin{array}{rcl}
  \Bdnf {\BL \leaf} &=& \{(\varnothing,\varnothing,\leaf)\}\\
  \Bdnf {\BN{\atom}{\Bp}{\Bm}} &=& \{ (A_p\cup\{\atom\},A_n,\leaf) \alt (A_p, A_n, \leaf)\in\Bdnf{\Bp} \}~\cup~\\
  &&\{ (A_p,A_n\cup\{\atom\},\leaf) \alt (A_p, A_n, \leaf)\in\Bdnf{\Bm} \}
\end{array}\]
\end{definition}

\begin{wrapfigure}{l}{.5\textwidth}
  \centering
    \begin{forest}
    for tree={circle, draw, 
              minimum size=1.5em,
              inner sep=1pt}
    [$\atom_1$
      [$\atom_2$
        [$\atom_3$
          [$\LE$]
          [$\LA$]
        ]
        [$\LA$]
      ]
      [$\atom_3$
        [$\LE$]
        [$\LA$]
      ]
    ]
  \end{forest}
  \captionof{figure}{\centering BDT whose interpretation is\\$\Bint{\atom_1}\land(\Bint{\atom_2}\land\neg\Bint{\atom_3} \vee \neg\Bint{\atom_2})\vee\neg\Bint{\atom_1}\land\neg\Bint{\atom_3}$\label{fig:bdd}}
\end{wrapfigure}

The knowledgeable reader may recognize our BDTs as variants of Reduced Ordered
Binary Decision Diagrams (ROBDD) \cite{ROBDD}. There are however two key
distinctions between the two formalisms. The first one is that for BDTs, leaf
nodes may be arbitrary elements of a lattice (that can be mapped to the lattice
of types). The second one is that we do not require BDTs to be DAGs, that is, we
do not enforce that structurally equal subtrees are shared in memory. While it
is possible to enforce this property (e.g. by performing hash-consing at
construction time), it has two drawbacks. The first one is
that, in typical use,
it is seldom the case that identical (and sufficiently large) subtrees are built
to offset the cost of hash-consing. The second is that in our setting, we may
have semantically equivalent subtrees which are not syntactically equivalent and
thus will not be compacted by hash-consing. We set these considerations aside
for now, and will revisit them in Section~\ref{sec:eval}.

\subsection{Set-Theoretic Operations}\label{sec:bdd-st}

Implementing set-theoretic operations on BDTs is straightforward.
For negation, it suffices to recursively traverse the BDT and negate the leaves:
\begin{definition}
  The negation of a BDT can be computed by negating its leaves:
  \[\begin{array}{rclcrcl}
  \neg\BL{\leaf} &=& \BL{\neg\leaf}&\hspace*{10mm}&
  \neg\BN{\atom}{\Bp}{\Bm} &=& \BN{\atom}{\neg\Bp}{\neg\Bm}
  \end{array}\]
\end{definition}
For binary set-theoretic operations, their definitions can be given in a generic way by propagating
the operations down the leaves, following the $\tleq$ order over atoms.
\begin{definition}
Let $\stbinop \in\{ \wedge,\vee,\setminus \}$ be a binary set-theoretic operation.
Let $\B_1$ and $\B_2$ be two BDTs. $\B_1 \stbinop \B_2$ can be computed inductively as follows:
\begin{align*}
  \BL{\leaf_1}\stbinop \BL{\leaf_2} &= \BL{\leaf_1\stbinop\leaf_2}\\
  \BN{\atom}{\Bp}{\Bm}\stbinop \BL{\leaf} &= \BN{\atom}{\Bp\stbinop \BL{\leaf}}{\Bm\stbinop \BL{\leaf}}\\
  \BL{\leaf}\stbinop\BN{\atom}{\Bp}{\Bm} &= \BN{\atom}{\BL{\leaf}\stbinop\Bp}{\BL{\leaf}\stbinop\Bm}\\[1mm]
  \BN{\atom_1}{\Bp_1}{\Bm_1}\stbinop\BN{\atom_2}{\Bp_2}{\Bm_2} &=
\begin{dcases}
    \BN{\atom_1}{\Bp_1\stbinop\Bp_2}{\Bm_1\stbinop\Bm_2} & \text{if } \atom_1=\atom_2\\
    \BN{\atom_1}{\Bp_1\stbinop\BN{\atom_2}{\Bp_2}{\Bm_2}}{\Bm_1\stbinop\BN{\atom_2}{\Bp_2}{\Bm_2}} & \text{if } \atom_1\tleq\atom_2\\
    \BN{\atom_2}{\BN{\atom_1}{\Bp_1}{\Bm_1}\stbinop\Bp_2}{\BN{\atom_1}{\Bp_1}{\Bm_1}\stbinop\Bm_2} & \text{if } \atom_2\tleq\atom_1
  \end{dcases}
\end{align*}
\end{definition}


Finally, we define a map operation on BDTs (it will be used later to implement type substitutions).
\begin{definition}\label{def:bext}
Let $f:\Atoms\cup\Leaves\to \BDTs{\Atoms}{\Leaves}$ be a function
from atoms and leaves to BDTs. We define the extension
$\Bext f : \BDTs{\Atoms}{\Leaves}\to\BDTs{\Atoms}{\Leaves}$ of $f$ inductively, as follows:
  \[\begin{array}{rclcrcl}
  \Bext f(\BL{\leaf}) &=& f(\leaf)&\hspace*{10mm}&
  \Bext f(\BN{\atom}{\Bp}{\Bm}) &=& (f(\atom)\land \Bext f(\Bp)) \vee (\neg f(\atom) \land \Bext f(\Bm))
  \end{array}\]
\end{definition}

\subsection{Semantic Simplification}\label{sec:bdd-simpl}

Unlike BDDs traditionally used for representing Boolean formulas whose atoms are
propositional variables, our BDTs may have more complex atoms (e.g. arrows, products).
Two atoms may be semantically equivalent without being syntactically
equivalent, or more generally, they may be correlated by the subtyping relation.
These subtyping relations between the atoms are not captured by the total order $\tleq$
we use for ordering the nodes of our BDTs, as this total order is purely syntactic.
Hence, we define a simplification operation that removes nodes that are (semantically) redundant from a BDT.
Note that this simplification operation does not try to reorder nodes or to change the atom
they are labeled with. As such, it does not guarantee minimality of the
resulting BDT, as a smaller semantically equivalent BDT that uses different
atoms or a different order may exist.


We define our BDT semantic simplification operation inductively as follows:
\[\begin{array}{ccccc}
  \Bsimpl{t}{\BL{\leaf}} &= \BL{\leaf} & \hspace*{10mm} &
  \Bsimpl{t}{\BN\atom\Bp\Bm}  &= {\begin{cases}
    {\Bp}' & \text{ if } t\land\Bint{{\Bp}'}\simeq t\land \Bint{\B'}\\
    {\Bm}' & \text{ if } t\land{\Bint{{\Bm}'}}\simeq t\land \Bint{\B'}\\
    \B' & \text{ otherwise}
  \end{cases}}
\end{array}\]
where ${\Bp}'=\Bsimpl{t\land\atom}{\Bp}$,\;\; ${\Bm}'=\Bsimpl{t\land \neg \atom}{\Bm}$,\;\; and $\B'=\BN\atom{{\Bp}'}{{\Bm}'}$.
The simplified form $\Bsimple\B$ of a BDT $\B$ is then defined as $\Bsimple\B=\Bsimpl\Any\B$.

In essence, the type $t$ in $\Bsimpl{t}{\B}$ represents the context of $\B$,
that is, the conjunction of positive and negative atoms traversed from the root to $\B$.
For each node $\B=\BN\atom\Bp\Bm$, if we note $t$ its context,
the simplification procedure tests whether $t\wedge \Bint{\B} \simeq t\wedge \Bint{\Bp}$.
If that is the case, it means that the atom and right subtree are redundant and
that $B$ can be replaced by (the simplified version of) $\Bp$. Otherwise, the same test is done for
$t\wedge \Bint{\B} \simeq t\wedge \Bint{\Bm}$, which allows one to replace $B$ by (the simplified version of)
its right subtree. If both tests fail, the node is kept and its subtrees are recursively simplified.

Note that simplifying a BDT may be expensive as it requires deciding several
semantic equivalences, each encoded as two subtyping tests. The simplification
procedure is correct:

\begin{proposition}[Correctness]\label{prop:correctness}
  For any BDT $\B$, we have $\Bint{\Bsimple\B} \simeq \Bint{\B}$.
\end{proposition}

More interestingly, while the simplification procedure does not ensure minimality, it enjoys two desirable properties.
First, the negation of a simplified type is itself simplified:
\begin{proposition}[Negation]\label{prop:bsimple_neg}
  For any BDT $\B$, we have $\neg\Bsimple{\B}=\Bsimple{\neg\B}$ (where $=$ is syntactic equality).
\end{proposition}
Second, the simplification is sufficient to always reduce the empty type to the trivial BDT:
\begin{proposition}[Emptiness]\label{prop:bsimple_empty}
  For any BDT $\B$, we have $\Bint{\B}\simeq\Empty \Leftrightarrow \Bsimple{\B}=\BE$.
\end{proposition}

This guarantees that, if a BDT is simplified, then checking its semantic emptiness
is equivalent to checking its syntactic emptiness.
Both Propositions~\ref{prop:bsimple_neg} and \ref{prop:bsimple_empty} will be exploited
by our representation of set-theoretic types to improve performance (cf. Section~\ref{sec:node_simpl}).
Proofs of these propositions are available in Appendix~\ref{app:proofs}.

\section{Internal Representation of Types}\label{sec:nodes}

In this section, we propose a structure to represent set-theoretic types using BDTs (cf. Section~\ref{sec:bdd}).
This structure allows building types using different operations:
\begin{itemize}
  \item type constructors (e.g. $\times$, $\to$, constants) and set-theoretic connectives ($\land$, $\lor$, $\neg$),
  \item application of a type substitution $\{\alpha_i\leadsto t_i\}_{i\in I}$ on a type $t$,
  \item construction of recursive types from a set of equations $\{\alpha_i = t_i\}_{i\in I}$.
\end{itemize}

For simplicity, we only consider the constant, arrow, and product type constructors.
Our structure is modular, so adding other type constructors is straightforward and is discussed
in Section~\ref{sec:extensions}.

\subsection{Overall Structure}

Set-theoretic types are defined coinductively and, thus, may be infinite trees.
However, the regularity property ensures that the number of distinct subtrees is finite:
we can thus represent a type as a graph that may contain cycles.
We build this graph such that each subtree whose parent is a type constructor ($\times$ or $\to$)
is represented by a node. A node thus defines the top-level structure of a subtree (i.e. the Boolean combination
of constructors and type variables composing it), and references other nodes in the graph to describe
the subtrees appearing inside a type constructor.
The contractivity property of types ensures no such node definition consists in an infinite union or intersection.
This graph structure is illustrated in Figure~\ref{fig:nodes}.


\begin{figure}[h]
\begin{minipage}{.33\textwidth}
  \centering
  \ADJUST{5}
\scalebox{0.9}{
\begin{tikzpicture}[
    node style/.style={
        draw,
        rounded corners,
        inner sep=6pt,
        minimum width=1.5cm,
        minimum height=1cm,
        font=\small
    },
    >=Stealth,
    thick
    ]
    \node[node style] (N1) {$N_1: (N_2 \times N_1) \vee \Nil$};
    \node[node style, below=of N1] (N2) {$N_2:\alpha \land (N_3 \to N_4)$};
    \node[node style, below=of N2.west,xshift=0.5cm] (N4) {$N_3: \Int$};
    \node[node style, below=of N2.east,xshift=-0.5cm] (N3) {$N_4: \Bool$};
    \draw[->] (N1) -- (N2);
    \draw[->] (N1) edge[loop above] (N1);
    \draw[->] (N2) -- (N3);
    \draw[->] (N2) -- (N4);
\end{tikzpicture}
}
  \captionof{figure}{\centering Graph for the type\\\hspace*{-1cm}$\mu X.\;(\pair{(\alpha\land(\Int\to\Bool))}{X})\vee\Nil$\hspace*{-1cm}\label{fig:nodes}}
\end{minipage}
~
\begin{minipage}{.65\textwidth}
  \centering
  \hspace*{-5mm}
  \scalebox{0.85}{








\begin{tikzpicture}[
  box/.style={
    draw,
    rounded corners,
    thick,
    minimum width=30mm,
    minimum height=12mm,
    inner sep=5pt,
    align=center,
    font=\small
  }
]

\node[box, fill=blue!5, minimum width=9.2cm] (node) {%
  \textbf{Node $\Node$}\\[2pt]
  represents a set-theoretic type (identifier, definition $\VDescr$)
};

\node[box, fill=green!5, minimum width=9.2cm, below=2pt of node] (poly) {%
  \textbf{Polymorphic descriptor $\VDescr$}\\[2pt]
  combines \emph{monomorphic descriptors} with \emph{type variables} ($\vee$ $\wedge$ $\neg$ $\alpha$ $\Descr$)
};

\node[box, fill=yellow!10, minimum width=9.2cm, below=2pt of poly] (mono) {%
  \textbf{Monomorphic descriptor $\Descr$}\\[2pt]
  represents a disjoint union of \emph{components} ($A\vee C \vee P$)
};

\node[box, fill=orange!10, below=2pt of mono, xshift=-31mm] (comp1) {%
  \textbf{Arrow component}\\[2pt]
  $\vee$ $\wedge$ $\neg$ $(\Node\to\Node)$
};
\node[box, fill=orange!10, below=2pt of mono] (comp2) {%
  \textbf{Const. component}\\[2pt]
  $\vee$ $\wedge$ $\neg$ $c$
};
\node[box, fill=orange!10, below=2pt of mono, xshift=31mm] (comp3) {%
  \textbf{Prod. component}\\[2pt]
  $\vee$ $\wedge$ $\neg$ $(\pair\Node\Node)$
};

\node[box, fill=red!10, below=2pt of comp1] (atom1) {%
  \textbf{Arrow atom}\\[2pt]
  $\Node\to\Node$
};
\node[box, fill=red!10, below=2pt of comp2] (atom2) {%
  \textbf{Constant atom}\\[2pt]
  $c$
};
\node[box, fill=red!10, below=2pt of comp3] (atom3) {%
  \textbf{Product atom}\\[2pt]
  $\pair\Node\Node$
};

\node[draw, rounded corners, thick, fit=(node)(poly)(mono)(comp1)(comp2)(comp3)(atom1)(atom2)(atom3), inner sep=3pt] {};

\draw[->, thick, bend left=45]
  ([xshift=5mm,yshift=5mm]atom1.west |- atom3.south) to
  node[left, align=center, font=\small, xshift=6pt,rotate=90,anchor=center]{references}
  ([xshift=5mm]node.west);
\draw[->, thick, bend right=45]
  ([xshift=-5mm,yshift=5mm]atom3.east |- atom1.south) to
  node[right, align=center, font=\small, xshift=-7pt,rotate=90,anchor=center]{references}
  ([xshift=-5mm]node.east);
\end{tikzpicture}
}
\ADJUST{5}
  \captionof{figure}{\centering Structure of a node (lower layers are used by upper layers)\label{fig:structure}}
\end{minipage}%
\end{figure}

Figure~\ref{fig:structure} illustrates the internal structure of each node of the graph.
Each node can be decomposed into several layers. At the bottom, we have the different kinds of \emph{atoms}
(i.e. type constructors): products $\times$, arrows $\to$, and constants.
Atoms of each kind are used inside \emph{components} that represent a Boolean combination of atoms.
Those different kinds of components are all disjoint (products, arrows and constants have disjoint interpretations):
their disjoint union is represented by a \emph{monomorphic descriptor}.
Finally, a \emph{polymorphic descriptor} expresses a Boolean combination of type variables
and monomorphic descriptors.
Concrete type definitions of all these structures are given in Type~Definition~\ref{typedef:stt}, in an OCaml-like syntax.

\begin{figure}[h]
\begin{typedef}[Set-Theoretic Types]
~\\[5mm]
\begin{minipage}{0.98\textwidth}
\begin{stml}[numbers=left,firstnumber=1,xleftmargin=4mm,framexleftmargin=3mm,numbersep=1mm]
(* Binary Decision Trees (Section $\ref{sec:bdd}$) *)
type ('atom,'leaf) bdt = Node of 'atom * t * t  |  Leaf of 'leaf
(* Atoms (Section $\ref{sec:atoms}$) *)
type const_atom        = { const_id: int ; const_name: string }
and  arrow_atom        = { dom: node ; codom: node }
and  product_atom      = { pi1: node ; pi2: node }
(* Components (Section $\ref{sec:components}$) *)
and  const_comp        = (const_atom,   bool) bdt
and  arrow_comp        = (arrow_atom,   bool) bdt
and  product_comp      = (product_atom, bool) bdt
(* Monomorphic descriptors (Section $\ref{sec:mono_descr}$) *)
and  mono_descr        =
     { const: const_comp ; prod: product_comp ; arrow: arrow_comp }
(* Polymorphic descriptors (Section $\ref{sec:poly_descr}$) *)
and  type_var          = { var_id: int ; var_name: string }
and  poly_descr        = (type_var, mono_descr) bdt
(* Nodes (Section $\ref{sec:nodes_def}$) *)
and  node              = { id: int ; def: poly_descr }
\end{stml}
\end{minipage}
\label{typedef:stt}
\end{typedef}
\end{figure}

In the following, nodes are ranged over by $\Node$.
We assume we have a total order $\tleq$ on nodes (it can be any total order, unrelated to the subtyping
relation over types).

\begin{definition}
  A \emph{node substitution} $\ns$ is a function from nodes to nodes
  which is the identity everywhere except for a finite set of nodes,
  called its domain and denoted by $\dom{\ns}$.
\end{definition}

\subsection{Atoms}\label{sec:atoms}

An atom $\atom$ represents an instance of a type constructor.
For an atom $\atom$ of any kind (constant, product, arrow), we define its dependencies $\deps{\atom}$ and the application $\atom\ns$ of a node substitution $\ns$ to $\atom$.

\paragraph{Arrow} An arrow atom is a term $\Node_1\to\Node_2$.
Intuitively, it represents the type $t_1\to t_2$ where $t_1$ (resp. $t_2$)
is the type associated with $\Node_1$ (resp. $\Node_2$).
\[\begin{array}{rclcrcl}
  \deps{\Node_1\to\Node_2} &=& \{ \Node_1, \Node_2 \} & \hspace*{15mm} &
  (\Node_1\to\Node_2)\ns &=& \ns(\Node_1)\to\ns(\Node_2)
\end{array}\]

\paragraph{Product} A product atom is a term $\pair{\Node_1}{\Node_2}$.
Intuitively, it represents the type $\pair{t_1}{t_2}$ where $t_1$ (resp. $t_2$)
is the type associated with $\Node_1$ (resp. $\Node_2$).
\[\begin{array}{rclcrcl}
  \deps{\pair{\Node_1}{\Node_2}} &=& \{ \Node_1, \Node_2 \} & \hspace*{15mm} &
  (\pair{\Node_1}{\Node_2})\ns &=& \pair{\ns(\Node_1)}{\ns(\Node_2)}
\end{array}\]

\paragraph{Constant} A constant atom is a label $c\in\Constants$ (e.g. $\true$, $\false$, $\nil$).
\[\begin{array}{rclcrcl}
  \deps{c} &=& \varnothing & \hspace*{15mm} &
  c\ns &=& c
\end{array}\]

We also extend the total order $\tleq$ over atoms (we can use the lexicographic order
for arrows and products, and an arbitrary order over constants).

\subsection{Components}\label{sec:components}

A component represents a Boolean combination of atoms of a given kind.
For simplicity, all kinds of components will be represented using BDTs.
In practice, though, simple components can be represented using a
more efficient, tailored representation (e.g. finite/cofinite sets for constants).

\begin{definition}
The BDTs we use for components use \emph{Boolean leaves}, defined as follows:
  \begin{displaymath}
  \begin{array}{lrrl}
    \textbf{Boolean Leaves} &\bleaf &::=& \LE \alt \LA
  \end{array}
\end{displaymath}
We define the interpretation of a Boolean leaf as follows: $\Bint \LE = \Empty$, $\Bint \LA = \Any$.
Set-theoretic operations ($\land$, $\lor$, $\neg$) on Boolean leaves are defined accordingly.
\end{definition}

\begin{definition}
  A component $\C$ is a BDT whose leaves are Boolean leaves,
  and whose atoms are one kind of atoms defined above (arrows, products, constants).
\end{definition}

Components inherit the operations defined on BDTs (cf. Section~\ref{sec:bdd}):
$\wedge$, $\vee$, $\neg$, $\Batoms{\C}$, $\Bint{\C}$, $\Bdnf{\C}$.
For each component $\C$ and node substitution $\ns$, we define $\deps{\C}$ and $\C\ns$ as follows:
\[\begin{array}{rclcrcl}
  \deps{\C} &=& \textstyle\bigcup_{\atom\in\Batoms{\C}}\deps\atom &\hspace*{15mm}&
  \C\ns &=& \Bext f (\C)
\end{array}\]
where $\Bext f$ is the extension on BDTs (cf. Definition~\ref{def:bext}) of the following function $f$:
\[\begin{array}{rclcrcl}
  f(\bleaf) &=& \BL\bleaf &\hspace*{15mm}&
  f(\atom) &=& \BS{\atom\ns}
\end{array}\]

We also extend the total order $\tleq$ over components (we can choose any syntactic total order over
the structure of the underlying BDT).

\subsection{Monomorphic Descriptors}\label{sec:mono_descr}

A monomorphic descriptor $\Descr$ represents a disjoint union of components.
\begin{definition}
A \emph{monomorphic descriptor} is a record $\{ \dconsts ; \dpairs ; \darrows \}$,
where $\dconsts$ is a component for constants,
$\dpairs$ is a component for products, and $\darrows$ is a component for arrows.
\end{definition}

As these components have disjoint interpretations, all the set-theoretic operations ($\neg$, $\vee$, $\wedge$)
as well as node substitutions can be performed component-wise.
The set of dependencies $\deps\Descr$ of a descriptor $\Descr$ is the union of the dependencies of its components.
We extend the total order $\tleq$ over monomorphic descriptors (we can use the lexicographic order
over the different components).


\subsection{Polymorphic Descriptors}\label{sec:poly_descr}

\begin{definition}
A polymorphic descriptor $\VDescr$ is a BDT whose leaves are monomorphic descriptors,
and whose atoms are type variables (we fix an arbitrary total order $\tleq$ over type variables).
\end{definition}

Polymorphic descriptors inherit the operations defined on BDTs (cf. Section~\ref{sec:bdd}):
$\wedge$, $\vee$, $\neg$, $\Batoms{\VDescr}$, $\Bleaves{\VDescr}$, $\Bint{\VDescr}$, $\Bdnf{\VDescr}$.
We define $\deps{\VDescr}$ and $\VDescr\ns$ (where $\ns$ is a node substitution) as follows:
\[\begin{array}{rclcrcl}
  \deps{\VDescr} &=& \textstyle\bigcup_{\Descr\in\Bleaves{\VDescr}}\deps\Descr &\hspace*{15mm}&
  \VDescr\ns &=& \Bext f (\VDescr)
\end{array}\]
where $\Bext f$ is the extension on BDTs (cf. Definition~\ref{def:bext}) of the following function $f$:
\[\begin{array}{rclcrcl}
  f(\Descr) &=& \BL{\Descr\ns} &\hspace*{15mm}&
  f(\alpha) &=& \BS{\alpha}
\end{array}\]

\begin{definition}
  A \emph{definition mapping} $\ds$ is a mapping from type variables
  to polymorphic descriptors. Its domain is denoted by $\dom{\ds}$.
\end{definition}

The application of a mapping $\ds$ on a polymorphic descriptor $\VDescr$
is defined as $\VDescr\ds=\Bext f (\VDescr)$,
where $\Bext f$ is the extension on BDTs (cf. Definition~\ref{def:bext}) of the following function $f$:
\[\begin{array}{rclcrcl}
  f(\Descr) &=& \BL{\Descr} &\hspace*{15mm}&
  f(\alpha) &=& \begin{dcases}
    \ds(\alpha)&\text{if }\alpha\in\dom\ds\\
    \BS{\alpha}&\text{otherwise}
  \end{dcases}
\end{array}\]

We extend the total order $\tleq$ over polymorphic descriptors (we can choose any syntactic total order over
the structure of the underlying BDT).

\subsection{Nodes}\label{sec:nodes_def}

\begin{definition}
A node is represented by a record $\{ \tid ; \tdef \}$
where $\tid$ is a unique identifier (e.g. an integer),
and $\tdef$ is a polymorphic descriptor.
\end{definition}

The total order $\tleq$ on nodes only depends on their $\tid$ field.
We write $\ntype{d}$ to denote a node with the definition $d$ and a fresh identifier.
Set-theoretic operations on nodes are defined as follows:
\[\begin{array}{ccc}
  \neg \Node = \ntype{\neg(\Node.\tdef)} &
  \Node_1 \land \Node_2 = \ntype{\Node_1.\tdef \land \Node_2.\tdef} &
  \Node_1 \lor \Node_2 = \ntype{\Node_1.\tdef \lor \Node_2.\tdef}
\end{array}\]

The set of top-level type variables of a node $\Node$ is defined as $\tlvars \Node = \Batoms{\Node.\tdef}$.
The set of dependencies of a node $\Node$, $\deps \Node$, is the smallest set of nodes $S$ such that
$\Node\in S$, and $\forall \Node'\in S.\; \deps {\Node'.\tdef} \subseteq S $.
The set of type variables of a node $\Node$ is defined as $\vars \Node = \bigcup_{\Node'\in\deps \Node} \tlvars{\Node'}$.

\begin{definition}
  A \emph{type substitution} $\s$ is a function from type variables to nodes
  which is the identity everywhere except for a finite set of type variables,
  denoted by $\dom{\s}$.
\end{definition}

The application $\Node\s$ of a type substitution $\s$ on a node $\Node$ is the node $\ns(\Node)$ where:
\begin{itemize}
  \item $\ns$ is a node substitution $\{ \Node_i \leadsto \Node_i' \}_{i\in I}$ where $\{\Node_i\}_{i\in I}=\deps\Node$, and
  \item for every $i\in I$, $\Node_i'=\ntype{((\Node_i.\tdef)\ns)\ds}$, and
  \item $\ds$ is the definition mapping $\{\alpha\leadsto \s(\alpha).\tdef\}_{\alpha\in\dom\s}$
\end{itemize}
Intuitively, the connected component of the node is copied ($\ns$ associates each node to its copy),
and the definition mapping $\ds$ is applied to the definition of each copied node.

Note that this definition is recursive: the definition of the node
$\Node_i'$ depends on $\ns$, which itself involves $\Node_i'$. In SSTT,
the field $\Node.\tdef$ is mutable, making it possible to generate a fresh node and only
set its definition later (it also allows us to simplify the definition of a node
after it is created, as it will be discussed in Section~\ref{sec:node_simpl}).
Alternatively, these equations can be implemented
by storing node definitions in a separate data-structure (for instance
a dictionary mapping node identifiers to their definitions), or
in a lazy language (e.g. Haskell) using recursive definitions.

\subsection{Construction of Recursive Types}

Nodes can be constructed from atoms and combined using set-theoretic operations.
However, we do not have a way to construct recursive types yet.
We propose here a method for building a recursive type from a set of equations.

Let $\Node$ be a node and $E$ be a set of equations, each equation being
a pair $(\alpha,\Node_\alpha)$ corresponding to the equation $\alpha = \Node_\alpha$.
Let $S=\bigcup_{(\alpha,\Node_\alpha)\in E}\deps{\Node_\alpha}$.
Using a topological sort algorithm, we order $S$ into an ordered set $\{\Node_i\}_{i\in 1\isep n}$ such that
$\forall i\in 1\isep n.\; \forall\alpha\in\tlvars{\Node_i}.\; \forall j\in i\isep n.\; (\alpha,\Node_j)\notin E$.

In other words, our order must guarantee that for every binding $(\alpha,\Node_\alpha)$ of our set of equations $E$,
$\Node_\alpha$ is smaller than all the nodes that depend on $\alpha$ at top-level.
If no such ordering exists, then it means that the set of equations is not contractive
(it contains dependency cycles that do not pass under a type constructor):
there may be no solution or infinitely many solutions that satisfy the set of equations,
and thus we reject it.

Otherwise, for every $i\in 1\isep n$, we define a node $\Node'_i = \ntype{((\Node_i.\tdef)\ns)\ds_i}$, where:
\begin{itemize}
  \item $\ns$ is the node substitution $\{ \Node_j \leadsto \Node_j' \}_{j\in 1\isep n}$, and
  \item $\ds_i$ is the definition mapping $\{\alpha\leadsto \Node_j'.\tdef \alt j\in 1\isep (i-1), \alpha\in\Vars \text{ s.t. } (\alpha,\Node_j)\in E \}$
\end{itemize}

As for type substitutions, this definition is recursive and should be implemented using similar techniques.
The function $\solve{\Node}{E}$ returns a node $\Node\s$
where $\s=\{\alpha\leadsto \Node'_i\alt (\alpha,\Node_i)\in E\}$. If $E$ is not contractive,
$\solve{\Node}{E}$ is not defined.

\subsection{Node Simplification}\label{sec:node_simpl}

Depending on how they are constructed, our nodes may contain redundant atoms in their internal representation.
For instance, consider the following sequence of operations:
$(i)$ the type $\Bool\to\Bool$ is constructed, $(ii)$ it is intersected with $\True\to\False$ (where $\True$ and $\False$ are singleton types that are subtypes of $\Bool$), and $(iii)$ it is intersected with $\False\to\True$.
The type we obtain is semantically equivalent to $(\True\to\False)\land(\False\to\True)$,
but its internal representation will still contain the atom $\Bool\to\Bool$.
Redundancy can also accumulate after applying multiple successive substitutions on a type,
which typically happens when implementing a type inference algorithm based on tallying such as \cite{popl24}.

The BDT simplification procedure $\Bsimple{.}$ defined in Section~\ref{sec:bdd-simpl} can be used to avoid
the accumulation of redundant atoms. In SSTT, we implemented a systematic simplification of types:
a node is automatically simplified after being constructed by a $\land$, $\lor$, a substitution, or a $\solvef$ operation.
According to Proposition~\ref{prop:bsimple_neg},
a type does not need to be simplified again after a $\neg$ operation.
Keeping node definitions in a simplified form allows us to decide emptiness
just by checking if the definition of the node is syntactically the empty BDT (cf. Proposition~\ref{prop:bsimple_empty}).
This means that, though the subtyping algorithm (Section~\ref{sec:subtyping-tallying})
is used to simplify the definition of a node initially, subsequent emptiness checks are constant time.
The time overhead induced by a systematic simplification of types is measured and discussed in Section~\ref{sec:eval}.

\section{Type Operations}\label{sec:subtyping-tallying}

In the rest of this paper, we identify a type $t$ with a node $\Node$ that represents it.
Set-theoretic operations and substitutions on types correspond to the associated operations on nodes.

\subsection{Subtyping}\label{sec:subtyping}
As already explained, the subtyping
algorithm is really an emptiness test. Following \cite{polysub}, to test a type $t$ for emptiness one only has to:
\begin{itemize}
\item Compute $\textsf{DNF}(t)$ as in equation \eqref{eqdnf}
\item Disregard the top-level variables $\alpha_i^{\{\textsf{b},\textsf{p},\textsf{a}\}}$ and $\beta_j^{\{\textsf{b},\textsf{p},\textsf{a}\}}$
\item Test that all combinations of basic types, product types and arrow types are empty
\end{itemize}
The reason for the second condition is that, since the type must be empty
\emph{for all} possible substitutions, it must be in particular for those mapping positive
variables to $\Any$ and negative variables to $\Empty$. We detail a first
version of the emptiness test in Algorithm~\ref{code:subtyping1}.
In this code\footnote{We use an OCaml-like syntax to present algorithms.},
$\Node$ is the node of the type we are testing and $\Sigma$ is a set of nodes. Each
layer (in the sense of Figure~\ref{fig:structure}) handles a particular aspect of the
emptiness test. Recursive types are handled at the level of nodes, by
remembering which node has already been visited. If a node is encountered during
its own traversal, we consider it is empty, since a recursive type whose
emptiness depends only on itself, e.g. $\mu X.(\Any, X)$\footnote{The model of
semantic subtyping only allows for finite values.}, is empty. 

\begin{wrapfigure}{l}{0.5\textwidth}
\begin{algorithm}[Subtyping, non-optimized]
  ~
\begin{stml}[numbers=left,firstnumber=1,xleftmargin=4mm,framexleftmargin=3mm,numbersep=1mm]
let rec is_empty $\Node$ $\Sigma$ =
  if $\Node\in\Sigma$ then true
  else
    let $\Sigma'$ = $\Sigma\cup\{ \Node \}$ in
    let $\VDescr$ = $\Node$.def in
    $\forall \Descr\in\Bleaves{\VDescr}$,
      is_empty_const $\Descr$.const $\Sigma' ~\wedge$
      is_empty_prod $\Descr$.prod $\Sigma' ~\wedge$
      is_empty_arrow $\Descr$.arrow $\Sigma' ~$
\end{stml}
\label{code:subtyping1}
\end{algorithm}
\end{wrapfigure}

The first time a node is visited, it is recorded (l.~4), and its polymorphic descriptor $\VDescr$
(a BDT whose atoms are type variables) is inspected. At that point, it is
sufficient to inspect all the monomorphic descriptors at the leaves of the
BDT. Each such descriptor $\Descr$ is a disjoint union of BDTs of constants,
products, and arrows, for which the emptiness test may call \code{is\_empty}
again on nested nodes. Thus, the two base cases of this recursive algorithm are:
when a node has already been visited (l.~2), which will happen if the input type is recursive,
and when the definition of the node is trivial and none of the calls to
\code{is\_empty\_const}, \code{is\_empty\_prod}, and \code{is\_empty\_arrow} (l.~7--9) recurses.
These functions are detailed below.

\paragraph{Constant Component} Since constants are disjoint one from the other,
testing the emptiness of a BDT of
constants is done in constant time, by checking if it is
structurally equal to the leaf $\LE$.
\paragraph{Product Component}
Testing the emptiness of products is done by the \code{is\_empty\_prod}
function:
\begin{stml}[frame=none,numbers=none,firstnumber=1]
let is_empty_prod $p$ $\Sigma$ = $\textstyle \forall\, (\{\pair{P_1^i}{P_2^i}\}_{i\in I}, A_n,\LA) \in \Bdnf{p}.~
  \Psi_\textsf{prod}(\pair{\bigwedge_{i\in I} P_1^i}{\bigwedge_{i\in I} P_2^i}, A_n, \Sigma)$
\end{stml}
This function enumerates each conjunction of the DNF of $p$ whose leaf node is $\LA$.
For each one, the positive part of the intersection
is transformed into a single product, by pushing the intersection below the
product constructor. Note that the types $\bigwedge_{i\in I} P_1^i$ and $\bigwedge_{i\in I} P_2^i$
are constructed following the set-theoretic operations on nodes defined in Section~\ref{sec:nodes},
which themselves may rely on BDT operations (Section~\ref{sec:bdd}).
It then remains to test whether the negative part of the
intersection is enough to negate this single product, which is tested by the
$\Psi_\textsf{prod}$ function:
\begin{displaymath}
  \begin{array}{lcl}
      \Psi_\textsf{prod}(\pair{P_1}{P_2}, \varnothing, \Sigma) & = & (\code{is\_empty} ~P_1~ \Sigma) \vee (\code{is\_empty}~ P_2 ~\Sigma)\\[1mm]
      \Psi_\textsf{prod}(\pair{P_1}{P_2}, \{\pair{N_1}{N_2}\}\cup S, \Sigma) & = & (\code{is\_empty} ~P_1~ \Sigma) \vee (\code{is\_empty}~ P_2 ~\Sigma) ~\vee\\
      & &  (~ \Psi_\textsf{prod}(\pair{(P_1\setminus N_1)}{P_2}, S, \Sigma)
      \wedge~ \Psi_\textsf{prod}(\pair{P_1}{(P_2\setminus N_2)}, S, \Sigma))\\
  \end{array}
\end{displaymath}
Note that in the worst case, the $\Psi_\textsf{prod}$ function may perform an
exponential number of calls to \code{is\_empty}, each Boolean connective
introducing a potential backtracking point. Crucially, even though new types are
generated (e.g. when pushing the intersection below the products or when
computing type differences), only a finite number of such new types are created
(see \cite[Section 3.5]{polysub}) which ensures the termination of
the algorithm (the number of all nodes collected in $\Sigma$ is finite).

\paragraph{Arrow Component} Testing the emptiness of arrows works similarly to
the case of products. The only difference is that intersection and arrow
constructors do not commute, and that a positive intersection of arrows is never
empty. We can apply these principles to test emptiness as follows.
\begin{stml}[frame=none,numbers=none,firstnumber=1]
let$\,\,$is_empty_arrow$\,\,a\,\,\Sigma\,=\,\textstyle \forall\, (A_p, A_n, \LA) \in \Bdnf{a}.\,\,
  \exists\, (N_1\to N_2)\in A_n.\,\,\Psi_\textsf{arrow}(N_1, \lnot N_2, A_p, \Sigma)$
\end{stml}
For an intersection of positive arrows $A_p$ and negative arrows $A_n$, there
must be a single negative arrow $N_1\to N_2$, which negates the whole positive
intersection. For that to be true, it suffices that at least one arrow in $A_p$ is completely negated by the selected negative arrow (which makes the positive intersection empty).
This is done by the auxiliary $\Psi_\textsf{arrow}$ function defined as:
\begin{displaymath}
  \begin{array}{lcl}
      \Psi_\textsf{arrow}(N_1, T_2, \varnothing, \Sigma) & = & (\code{is\_empty} ~N_1~ \Sigma) \vee (\code{is\_empty}~ T_2 ~\Sigma)\\[1mm]
      \Psi_\textsf{arrow}(N_1, T_2, \{P_1\to P_2\}\cup A_p, \Sigma) & = &
      (\code{is\_empty} ~N_1~ \Sigma) \vee (\code{is\_empty}~ T_2 ~\Sigma)~\vee\\
      & &  (\Psi_\textsf{arrow}((N_1\setminus P_1), T_2, A_p, \Sigma)
      \wedge \Psi_\textsf{arrow}( N_1, (P_2\wedge T_2), A_p, \Sigma))\\
  \end{array}
\end{displaymath}
This function is similar to the one used for products, but takes a single negative arrow $N_1\to N_2$ and
checks that it is a supertype of the positive part (and thus, that their difference is empty). Notice
that $T_2$ is initially the negation of the co-domain $N_2$.

\paragraph{Recursive Types and Caching}
To explain how recursive types are handled, consider the example:
\begin{displaymath}
      X ~=~ (\pair{Y}{Y})\quad\quad
    Y ~=~ (\pair{\nil}{Y}) ~~\vee~~ (\pair{Z}{Z})~~\vee~~(\nil\to\nil)\quad\quad
    Z ~=~ \pair{Y}{\nil}
\end{displaymath}
where types are written as a set of recursive equations between nodes (and some basic types).
If one uses Algorithm~\ref{code:subtyping1} to
test the emptiness of $X$, the result is \code{false}. A relevant subset of the recursive
calls is given in Figure~\ref{fig:subtyping-calls}.
\begin{figure}[h]
\begin{minipage}{0.98\textwidth}
\begin{stml}[numbers=left,firstnumber=1,xleftmargin=5mm,framexleftmargin=5mm]
is_empty $X$ $\varnothing$
     is_empty_prod $\pair{Y}{Y}$ $\{X\}$
         is_empty $Y$ $\{X\}$
            is_empty_prod $(\pair{\nil}{Y})\vee(\pair{Z}{Z})$ $\{X, Y\}$
                (is_empty_const $\nil$ $\{X, Y\} ~\leadsto$ false
                $\vee$ is_empty $Y$ $\{X, Y\} ~\leadsto$ true)     (*recursive occurrence*)
              $\wedge$ is_empty_prod $\pair{Z}{Z}$  $\{X,Y,Z\} ~\leadsto$ true (*depends on ${\color{cadmiumgreen}Y}$*)
          $\wedge$ is_empty_arrow $(\nil\to\nil)$ $\{X, Y\} \leadsto$ false
       $\vee$ is_empty $Y$ $\{X\}$
\end{stml}
\end{minipage}
\caption{Recursive calls for testing the emptiness of $X$\label{fig:subtyping-calls}}
\end{figure}
To test the emptiness of $X$, we must test its definition,
since $X$ is not in $\Sigma$ (l.~2). Its definition is the product type
$\pair{Y}{Y}$, which is empty if either projection is empty.
Those are tested at l.~3 and l.~9. This illustrates an inefficiency of the
algorithm. The set $\Sigma$ is used to handle recursion, and not as a
cache, which means that several occurrences of the same node may be traversed several times.

While it is tempting to replace $\Sigma$ with a mutable map,
one must take special care in doing so.
Indeed, in this example, it would be recorded that:
\begin{itemize}
  \item at first, $Y$ is empty (initial recursive call, l.~3)
  \item during its traversal, $Z$ (which depends on $Y$) is found to be empty
  \item ultimately, $Y$ contains $\nil\to\nil$ and therefore is \emph{not} empty
\end{itemize}

It is not sufficient to update $\Sigma$ to record that $Y$ is non-empty.
One must also \emph{invalidate} the results stored for all nodes which used the \emph{wrong} initial assumption that $Y$ was empty. This improvement is
described in Algorithm~\ref{code:subtyping3}.

The algorithm maintains three data-structures:
$(i)$ the hash table \code{memo} (l. 1), $(ii)$ a reference to a persistent list, used as a stack (\code{stack}, l. 2), and $(iii)$ a hash table mapping nodes to stacks (\code{mem\_stack}, l. 3).
While \code{memo} plays the same purpose as in the previous algorithm, the
other two data-structures are used to track dependencies between nodes.
Assuming a node $\Node$ is not in \code{memo}:
\begin{itemize}[leftmargin=10pt]
  \item push it on \code{stack} (l.~11)
  \;\;$\bullet$ create an entry in \code{mem\_stack} (l.~12)
  \;\;$\bullet$ map it to \code{true} in \code{memo} (l.~13)
\end{itemize}
We then explore the descriptor of the node recursively.

\begin{wrapfigure}{l}{0.5\textwidth}
\begin{algorithm}[Subtyping, optimized]
~\label{code:subtyping3}
\begin{stml}[numbers=left,firstnumber=1,xleftmargin=4mm,framexleftmargin=3mm,numbersep=1mm]
let memo = H.create ()
let stack = ref []
let mem_stack = H.create ()
let rec is_empty $\Node$ =
  if $\Node\in$ memo then begin
   if $\Node \in$ mem_stack then
    H.add mem_stack $\Node$
    (!stack::H.find mem_stack $\Node$);
   H.find memo $\Node$
  end else begin
   stack := $\Node$ :: !stack;
   H.add mem_stack $\Node$ [];
   H.add memo $\Node$ true;
   let $\VDescr$ = $\Node$.def in
   let b = $\forall \Descr\in\Bleaves{\VDescr}$,
     is_empty_const $\Descr$.const $\wedge$
     is_empty_prod $\Descr$.prod $\wedge$
     is_empty_arrow $\Descr$.arrow in
   if not b then
    invalidate $\Node$ memo mem_stack;
   H.add memo $\Node$ b;
   H.remove mem_stack $\Node$;
   stack := List.tl !stack;
   b end
\end{stml}
\end{algorithm}
\end{wrapfigure}
\noindent
If a node $\Node$ is in \code{memo}, then:
\begin{itemize}[leftmargin=10pt]
  \item if an entry exists for $\Node$ in \code{mem\_stack}, record the current
  stack (l. 6--8)
  \item return the memoized value for $\Node$ (l. 9)
\end{itemize}
Lastly, after returning from a recursive call, if the node turns out to be
non-empty then:
\begin{itemize}[leftmargin=10pt]
  \item for each recorded stack for $\Node$, remove from $\code{memo}$ any node $X$ appearing on that stack, until
  $\Node$ is reached (done by function \code{invalidate}, l.~20)\footnote{Due to the monotonic nature of \code{is\_empty}, false entries (non-empty) do not need to be removed~\cite[Section~7.1.2]{AlainThesis}.}
  \item update \code{memo} with the result for $\Node$ (l.~21)
  \item remove $\Node$ from \code{mem\_stack} (l.~22)
  \item pop $\Node$ from the stack (l.~23)
  \item return the result (l.~24)
\end{itemize}

Whenever we encounter a node $\Node$ a second time during a recursive traversal, either:
\begin{itemize}[leftmargin=10pt]
  \item it is in a definitive state (it has no binding in \code{mem\_stack}), we
  can therefore directly return its emptiness state from the \code{memo} table;
  \item it is still in an ``unknown'' state (i.e. it has a binding in \code{mem\_stack}),
  and nodes that have been put on the stack during its inspection depend on the (possibly wrong) assumption that it is empty,
  therefore we must remember them to remove them from \code{memo} later on.
\end{itemize}

\subsection{Tallying}\label{sec:tallying}

We revisit the tallying algorithm formalized by~\cite{polyduce2}, which does not
perform well in practice. At its heart, the tallying algorithm is fairly simple.
A tallying problem $\exists\sigma.\;s\sigma\leq t\sigma$ is turned into finding all
substitutions $\sigma$ for which $(s\setminus t)\sigma$ is empty. The tallying
procedure recursively explores the DNF of $s\setminus t$, generating a set of
sets of constraints. When performed naively, this procedure may be
suboptimal for two reasons. First, it may keep unsatisfiable constraints during
constraint generation to later on discard them. Second, it may keep
\emph{redundant} constraints which yield correct, but useless substitutions. For
instance, in the set of substitutions $\{ \{ \alpha\mapsto \Int \}, \{
\alpha\mapsto \Any \} \}$, the second one is useless (although correct) since
the first one is more precise. As our evaluation shows (Section~\ref{sec:eval}),
generating such sets of sets of constraints naively is very costly to the point that
some simple tallying problems cannot be solved in reasonable time. We
therefore propose an alternative presentation which incorporates eager
simplifications of constraints and greatly reduces the number of returned
substitutions. We first introduce \emph{normalized constraint sets}.
\begin{definition}[Normalized constraint set]
  A normalized constraint set $C$ is a set of triples $\{(s_i,\alpha_i,t_i)\}_{i\in I}$
  (representing the constraint $s_i\leq\alpha_i\leq t_i$) where all $\alpha_i$ are distinct.
  The set $\{\alpha_i\}_{i\in I}$ is called the \emph{domain} of $C$ and is written $\dom C$.
  The empty constraint set is noted $\CSAny$.
  Lastly, we define the constraint associated with a variable $\alpha$ in
  a constraint $C$ by:
 \begin{displaymath}
    C(\alpha) ~=~ (s, t)~\;~ \text{if } (s,\alpha,t)\in C \hspace*{2cm}
    C(\alpha) ~=~ (\Empty, \Any)~\;~ \text{otherwise}
 \end{displaymath}
\end{definition}
We now define
operations on normalized constraint sets. These are parameterized by
a set $\Delta$ of type variables the tallying
algorithm is allowed to instantiate.
\begin{definition}[Intersection of normalized constraint sets]
  Let $C_1$ and $C_2$ be two normalized constraint sets and $\Delta$ a set of type variables.
  We define the \emph{intersection} $C_1\ccap^\Delta C_2$ as follows.
  Let $C = \{ (s_1\vee s_2,\alpha,t_1\wedge t_2) ~|~ \alpha\in\dom{C_1}\cup\dom{C_2}, (s_i, t_i)=C_i(\alpha) \text{ for }i=1..2 \}$.
  \begin{displaymath}
    \begin{array}{lclr}
      C_1\ccap^\Delta C_2& = &  C & \text{if} ~\forall (s,\alpha,t)\in C, \vars{s}\cup\vars{t}\subseteq\Delta\Rightarrow s\leq t\\
      C_1\ccap^\Delta C_2& = & \bullet& \text{otherwise}
    \end{array}
  \end{displaymath}
\end{definition}
Intuitively, the intersection of two constraint sets is only defined if it does
not create a trivially unsatisfiable constraint such as $(\Int,\alpha,\Bool)$, which can
never be satisfied.

We now want to define intersections and unions over sets of constraint sets, ranged over by $\CSS$---each constraint set $C\in\CSS$ representing a potential solution of our tallying instance.
In order to avoid a combinatorial explosion and to generate a minimal set of solutions, we want to eliminate redundant constraint sets from $\CSS$: in particular, there should not be two different
constraint sets $C_1,C_2\in\CSS$ such that $C_1$ \textit{subsumes} $C_2$.
A first attempt at defining constraint set subsumption is to compare constraints for each variable independently, yielding the following definition:
\begin{definition}[Subsumption of constraint sets]
  Let $C_1$ and $C_2$ be two normalized constraint sets. We say that $C_1$ subsumes $C_2$ (intuitively, $C_1$ is more restrictive than $C_2$), written $C_1\sqsubseteq C_2$,
  if and only if:
  \begin{displaymath}
    \forall (s, \alpha, t) \in C_2.\;\; s \leq s' \text{ and } t' \leq t \text{ where } (s',t')=C_1(\alpha)
  \end{displaymath}
\end{definition}
In other words, a set of constraints $C_1$ subsumes a set of constraints $C_2$
if $C_1$ gives better bounds (higher lower bounds and lower upper bounds) for
all variables of $C_2$. However, this definition of subsumption is not sufficient to avoid
generating redundant solutions. Consider for instance the following two constraint sets:
\[ C_1=\{(\Int, \alpha, \Any),\; (\Any, \beta, \Any)\} \;\;\;\;\not\sqsubseteq\;\;\;\; \{(\Int, \alpha, \beta),\; (\Int, \beta, \Any)\}=C_2 \]

These two constraint sets are not related by $\sqsubseteq$: $C_2(\alpha)$
strictly subsumes $C_1(\alpha)$, while $C_1(\beta)$ strictly subsumes
$C_2(\beta)$. However, we should have $C_1\sqsubseteq C_2$: while
the constraint $(\Int,\alpha,\Any)$ does not subsume $(\Int,\alpha,\beta)$ by
itself, it does when we consider it in a context where $\beta\simeq\Any$ (which
is enforced by the constraint $(\Any,\beta,\Any)\in C_1$). To account for these
indirect relations, we refine our definition of constraint set subsumption:
\begin{definition}[Subsumption of constraint sets, improved]\label{def:subsumption-improved}
  Let $C_1$ and $C_2$ be two normalized constraint sets. We say that $C_1$ subsumes $C_2$, written $C_1\sqsubseteq C_2$, if and only if:
  \begin{displaymath}
    \forall (s, \alpha, t) \in C_2.\;\; \tlb{s}{C_1} \leq s' \text{ and } t' \leq \tub{t}{C_1} \text{ where } (s',t')=C_1(\alpha)
  \end{displaymath}
  with $\tub{u}{C}$ and $\tlb{u}{C}$ respectively a weakening and strengthening operator, inductively defined as follows
  (with $\tub{u}{\varnothing}=\tlb{u}{\varnothing}=u$):
  \begin{align*}
    \tub{u}{\{(s,\alpha,t)\}\cup C} &\;=\; \tub{u\tsubsp{\alpha}{\alpha\vee s}\tsubsm{\alpha}{\alpha\land t}}{C}&\text{where $\forall\alpha'\in\dom C.\;\alpha\tleq \alpha'$}\\
    \tlb{u}{\{(s,\alpha,t)\}\cup C} &\;=\; \tlb{u\tsubsp{\alpha}{\alpha\land t}\tsubsm{\alpha}{\alpha\vee s}}{C}&\text{where $\forall\alpha'\in\dom C.\;\alpha\tleq \alpha'$}
  \end{align*}
  where $u\tsubsp{\alpha}{s}$ denotes the type $u$ where every positive top-level occurrence of $\alpha$ has been substituted by $s$,
  and $u\tsubsm{\alpha}{s}$ denotes the type $u$ where every negative top-level occurrence of $\alpha$ has been substituted by $s$
  (both operations can easily be implemented on the BDT representation of the polymorphic descriptor of $u$).
\end{definition}
The side condition $\forall\alpha'\in\dom C.\;\alpha\tleq \alpha'$ ensures that constraints are propagated following their dependency order:
the tallying algorithm preserves the invariant that any constraint $(s,\alpha,t)$ is such that $\tlvars s$ and $\tlvars t$ only
contain type variables that are greater than $\alpha$ by $\tleq$ (where $\tleq$ is the same syntactic total order on type variables,
atoms and nodes that we use since Section~\ref{sec:bdd}).
See Appendix~\ref{app:subsumption} for a formal proof that $C_1\sqsubseteq C_2$ implies every solution of $C_1$ satisfies $C_2$.

As an example, consider our two constraint sets from earlier, $C_1=\{(\Int, \alpha, \Any),\; (\Any, \beta, \Any)\}$
and $C_2=\{(\Int, \alpha, \beta),\; (\Int, \beta, \Any)\}$. This time, we do have $C_1 \sqsubseteq C_2$:
when checking the constraints on $\alpha$, we are comparing $C_1$'s constraint $(\Int, \alpha, \Any)$ against $(\tlb{\Int}{C_1}, \alpha, \tub{\beta}{C_1})=(\Int, \alpha, \Any)$.
\begin{definition}[Simplification of sets of constraint sets]
The simplification of a set $\CSS$ of constraint sets, written $\CSSsimp(\CSS)$, is defined as follows:
\begin{displaymath}
  \begin{array}{lclr}
    \CSSsimp(\CSS) & = & \CSSsimp'(\CSS, \varnothing) \\[1mm]
\CSSsimp'(\varnothing, \mathscr{D}) & = & \mathscr{D} &\\
\CSSsimp'( \{ C  \} \cup \CSS, \mathscr{D}) & = &
   \CSSsimp'(\CSS, \mathscr{D}) & \text{if } \exists C'\in \mathscr{D}, C'\sqsubseteq C\\
   \CSSsimp'( \{ C \} \cup \CSS, \mathscr{D}) & = & \CSSsimp'( \CSS, \{C\} \cup \mathscr{D} \setminus \mathscr{D}') &
     \text{where } \mathscr{D}' = \{ C' ~|~ C'\in \mathscr{D}, C\sqsubseteq C'\}\\
\end{array}
\end{displaymath}
\end{definition}
Simplified sets of constraint sets ensure that their elements are pairwise non-subsumable, each element denotes a different, incomparable substitution.
Maintaining simplified sets is a key ingredient to tame the complexity of the tallying operation.
\begin{definition}[Union and intersection of sets of constraint sets]
  Given two sets of normalized constraint sets $\CSS_1$ and $\CSS_2$ and a set of type variables $\Delta$,
  we define their intersection and union as:
  \begin{displaymath}
    \begin{array}{lclr}
      \CSS_1 \ccap^{\Delta} \CSS_2 & = & \CSSsimp(\{ C_1\ccap^{\Delta} C_2 ~|~ C_1\in \CSS_1, C_2\in \CSS_2,  C_1\ccap^{\Delta} C_2 \neq \bullet \}) & \text{(intersection)}\\
      \CSS_1 \ccup \CSS_2 & = & \CSSsimp(\CSS_1 \cup \CSS_2) & \text{(union)}\\
    \end{array}
  \end{displaymath}
\end{definition}
We can now review the tallying algorithm, Algorithm~\ref{code:tallying}, which consists of three steps. 
The 
function \code{normalize} transforms a subtyping constraint $N_1 \leq N_2$ into
a set of normalized constraint

\noindent\begin{wrapfigure}{l}{0.48\textwidth}
  \ADJUST{3}
  \begin{algorithm}[Tallying]
    ~\label{code:tallying}
\begin{framed}
\ADJUST{2}
\begin{stml}[numbers=left,frame=none,firstnumber=1,%
  xleftmargin=0mm,xrightmargin=0mm,%
  numbersep=1mm,
  framesep=0pt,%
  rulesep=0mm,%
  linewidth=1.05\textwidth]
let tally $\Delta$ $S$ =
  let$\,\CSS$=$\displaystyle\!\!\!\!\!\bigsqcap_{(\Node_1,\Node_2)\in S}\!\!\!\!\!\!\!^\Delta\,$normalize $\Delta\,\,(\Node_1\!\setminus\!\Node_2)\,$in
  let$\,\CSS'$=$\displaystyle\bigsqcup_{C\in \CSS}$propagate $\Delta$ $\CSAny$ $C$ $\varnothing$ in
  $\{$ solve $C$ $ \alt C\in\CSS' \}$
\end{stml}
\ADJUST{2}
\end{framed}
\end{algorithm}
\ADJUST{5}
\end{wrapfigure} 
 sets $\CSS$.
Like in the subtyping algorithm, the initial
subtyping constraint is encoded as an emptiness constraint for the type $N_1 \setminus N_2$.
Each normalized constraint set is then processed through the \code{propagate}
function, which eliminates unsatisfiable constraint sets and recursively
enriches satisfiable ones with new induced constraints: for every constraint $(s, \alpha, t)$ produced
by \code{normalize},
if either $s$ or $t$ contains type variables that are not in $\Delta$, then
$s\leq t$ may require new constraints. 

\noindent These new constraints are added by \code{propagate} to form
a final set of constraint sets $\CSS'$. Each constraint set in $\CSS'$ can then be turned into a set
of recursive type equations and solved, yielding a substitution.
The code of \code{normalize} and \code{propagate} is given in Figure~\ref{code:tallyingdetail}.

\begin{figure}[t]
  \ADJUST{2}
\begin{framed}
\begin{stml}[frame=none,numbers=left,firstnumber=1,xleftmargin=5mm,aboveskip=-5pt,belowskip=-5pt]
let $M$ = H.create ()
let rec normalize $\Delta$ $\Node$ =
  if $\Node\in M$ then H.find $M$ $\Node$
  else begin
    H.add $M~\Node~\{\CSAny\}$;
    let $\CSS$ = $\displaystyle\hspace*{-1cm}\bigsqcap_{(A_p,A_n,\Descr)\in\Bdnf{\Node.\tdef}}\hspace*{-1.1cm}^\Delta$ normalize_summand $\Delta$ $A_p$ $A_n$ $\Descr$ in
    H.remove $M~\Node$; $\CSS$ end
and normalize_summand $\Delta$ $A_p$ $A_n$ $\Descr$ =
  let $A$ = $(A_p \cup A_n)\setminus\Delta$ in
  if $A\neq\varnothing$ then
   let $\alpha$ = min $A$ in (* Minimal type variable according to $\tleq$ *)
   if $\alpha\in A_p\,$then$\,\{\{(\Empty,\alpha,\neg(A_p\!\setminus\!\{\alpha\}\wedge A_n\wedge\Descr))\}\}\,$else$\,\{\{(A_p\wedge A_n\!\setminus\!\{\alpha\}\wedge\Descr,\alpha,\Any)\}\}$
  else (norm_const $\Delta$ $\Descr$.const) $\ccap^\Delta$ (norm_prod $\Delta$ $\Descr$.prod) $\ccap^\Delta$ ...
$\rule{0pt}{4mm}$let rec propagate $\Delta$ $C_0$ $C_1$ $\Sigma$ =
  if $C_1=\CSAny$ then $\{ C_0 \}$ else
    let $(\Node_1,\alpha,\Node_2)::C_1'$ = $C_1$ in let $\Node=\Node_1\setminus \Node_2$ in
    if $\Node\in\Sigma$ then propagate $\Delta$ ($C_0\ccap^\Delta\{(\Node_1,\alpha,\Node_2)\}$) $C_1'$ $\Sigma$
    else
      let $\CSS$ = $\{C_0 \ccap^\Delta C_1\}$ $\ccap^\Delta$ (normalize $\Delta$ $\Node$) in
      $\displaystyle\bigsqcup_{C\in\CSS}$ propagate $\Delta$ $\CSAny$ $C$ $(\Sigma\cup\{\Node\})$
\end{stml}
  \end{framed}
  \caption{The \code{normalize} and \code{propagate} functions}
  \label{code:tallyingdetail}
\ADJUST{2}
\end{figure}

\paragraph{Normalize} The implementation of \code{normalize} $\Delta$ $\Node$
follows the same structure as the emptiness test. It uses a hash table $M$
to map already visited nodes to their set of constraint sets (l.~1).
It plays a similar role as $\Sigma$ in the subtyping algorithm. Whenever
the emptiness test (which computes a Boolean) uses an ``and'' (resp. an ``or''),
the normalize function uses $\ccap$ (resp. $\ccup$). The first time a type is
visited (l.~5), the satisfiable constraint is added to the memoization table
$M$. Each conjunction of the DNF of the type is then explored to generate
individual sets of constraint sets which are intersected (l.~6). The node is
removed from the table $M$ and the resulting constraints returned.

The emptiness of a conjunction of the form $\textstyle\bigwedge_{\alpha\in
A_p}\alpha\land\textstyle\bigwedge_{\alpha\in A_n}\neg\alpha\land \Descr$ is expressed
by the function \code{normalize\_summand}. The function chooses the smallest
(according to the total order $\tleq$ on type variables) top-level variable that
is not in $\Delta$, and extract it from the summand, yielding a single
constraint on that variable denoting the emptiness of the summand (l.~10--12).
If there are no top-level variables, the function recursively explores the underlying constructors,
with a function that mimics the emptiness test on components (and recursively calls \code{normalize}).

\paragraph{Propagate} The function \code{propagate} $\Delta$ $C_0$ $C_1$ $\Sigma$ iteratively enriches the
constraint set $C_0 \ccap^\Delta C_1$ with the constraints induced by $C_1$ ($C_0$ contains the constraints already treated)
until it reaches a fixpoint ($\Sigma$ stores the induced constraints already visited). Note the interplay between
\code{normalize} and \code{propagate}. When \code{normalize} generates a single constraint for a top-level variable,
\code{propagate} calls \code{normalize} again on the bounds of that constraint.

\paragraph{Solve} The function \code{solve $C$}
represents the constraint set
$C$ as a principal type substitution. It makes use of the \code{build} function
(Section~\ref{sec:nodes}) to solve a set of equations. Each equation is
generated by turning a subtyping constraint $N_1\leq \alpha \leq N_2$ into an
equation $\alpha= (N_1\vee\alpha')\wedge N_2$, for a fresh $\alpha'$.

\section{Extensions}\label{sec:extensions}

In Section~\ref{sec:nodes} and Section~\ref{sec:subtyping-tallying}, we described
three core components: arrows, products, and constants.
These components are the most commonly used in set-theoretic type formalizations,
but a practical implementation of set-theoretic types also needs other core components
to model common programming structures (Section~\ref{sec:core-extensions}).
Additionally, these core components can be used as building blocks and combined
to define new components via encodings (Section~\ref{sec:encoding}).

\subsection{Core Extensions}\label{sec:core-extensions}

\paragraph{Intervals} The component for constants cannot be used
to represent all the individual integers, as the type $\Int$ would then need to be defined
as an infinite (or very large) union. Therefore, SSTT features
a component representing intervals as sorted lists of disjoint intervals.

\paragraph{Tuples} Products can be generalized to cover tuples of any arity.
For that, we can implement a generic tuple component parameterized by an arity $n$.
It is straightforward to extend the subtyping and tallying algorithms accordingly
(e.g. for $n=3$, a DNF of tuples $(t_1,t_2,t_3)$ is empty if and only if the corresponding DNF of pairs
$\pair{t_1}{(\pair{t_2}{t_3})}$ is empty).

\paragraph{Records} The theory of set-theoretic types features records that can be open or closed, and
fields that may be optionally absent. Record atoms are simply maps from field names to optional types, as well
as a Boolean indicating if the record is closed (its list of fields is exact) or open (it may have other, unknown fields).
In order to represent absent fields, we define optional types
as the pairs $(t,\abs)$ where $\abs$ is either $\top$ (the value may be absent) or $\bot$
(the value is not absent). An optional type $(t,\abs)$ is empty if and only if $t$ is empty and
$\abs=\bot$. Then, subtyping for records can be encoded as subtyping for tuples:
the record atoms composing a DNF can be turned into tuple atoms of optional types by
fixing an order over fields. For instance, the DNF $\crecord{\lbl_1\req t_1}\vee\orecord{\lbl_2\reqo t_2}$
can be encoded as $((t_1,\bot),(\Empty,\top),(\Empty,\top))\vee((\Any,\top),(t_2,\top),(\Any,\top))$
(the first component of the tuples represents the field $\lbl_1$,
the second component represents $\lbl_2$, and the last component represents the other fields).
Note that while our records can be open (extensible via subtyping), they do not feature row polymorphism, which requires substantial
modifications of the tallying procedure \cite{CP25,row-poly}.

\paragraph{Tagged Types} A tagged type $\T(t)$ intuitively represents the values $\{ \T(v) \alt v\in t \}$,
where $\T\in\Tags$ is a unary constructor of our language disjoint from other values
(in particular, two types $\T_1(t)$ and $\T_2(t)$ with $\T_1\neq \T_2$ are disjoint).
Tagged types have several uses:
$(i)$ they can be used to represent boxed values, and in conjunction with unions,
to implement algebraic data types where every constructor is disjoint from the others,
$(ii)$ they can be used as a building block to extend our type algebra with new type constructors
in a modular way using encodings (cf. Section~\ref{sec:encoding}), and
$(iii)$ in addition to the representation of boxed values,
they can be generalized to represent opaque data types with an invariant or covariant parameter,
as described in \cite{mlsem}. Extending the subtyping and tallying algorithms is straightforward, following
the inductive relations given in this last work.

\subsection{Encodings}\label{sec:encoding}

In addition to core extensions, new type constructors can be added by encoding them as combinations of
core components. This approach has two advantages: $(i)$ it allows the programmer to extend the type algebra in a modular way,
without having to modify the core components of the library,
and $(ii)$ it does not require any addition to the subtyping and tallying algorithms.
Designing such an extension requires answering these three questions:
\begin{enumerate}
  \item how to encode our new type constructors using the existing core components?
  \item how to recognize the encoding in a type?
  \item how to extract the parameters of our type constructors from their encoding?
\end{enumerate}
As long as we are only interested in subtyping and tallying, only the first point matters.
However, as soon as we want to be able to inspect our types (e.g. for pretty-printing, cf. Appendix~\ref{app:pp}),
it becomes necessary to recognize when a subtree of our type comes from an encoding,
and how to destruct this encoding.
In order to make our encodings easily recognizable in a type, each extension should be disjoint from the other values.
For instance, encoding the list type $\ListS{\alpha}$ as the type $X = {\Nil}~\vee~\pair{\alpha}{X}$ as we did in
Section~\ref{sec:intro} is ambiguous, as it becomes unclear whether $\pair\Int\Nil$ should be printed as a pair or as a list.
We solve this problem by using tagged types: each constructor of an extension is tagged by a distinct tag that ensures its disjointness
w.r.t. other types (e.g. list and pairs are kept apart) and can be used by the pretty-printer to defer printing to a specialized function,
provided by the extension.
\begin{table}[h]
  \centering
  \caption{Example of extensions and their encodings\label{tab:extensions}}
  \scalebox{0.85}{
\renewcommand{\arraystretch}{1.0} 
\begin{NiceTabular}{ |l|l|l| }
  \toprule
  Extension & Constructor & Encoding \\
  \midrule \rowcolor[rgb]{ .906,  .902,  .902}

                                 & Single character \code{'a'} & $\texttt{Chr}(i_a)$ where $i_a$ is the ASCII encoding of \code{'a'} \\\rowcolor[rgb]{ .906,  .902,  .902}
                                 & Char interval \code{'a'-'z'} & $\texttt{Chr}(i_a\isep i_z)$ \\\rowcolor[rgb]{ .906,  .902,  .902}
  \multirow{-3}{5.5em}{Characters} & Any (supertype) & $\texttt{Chr}(\texttt{int})$ \\

                                & Single string \code{"abc"} & $\texttt{Str}(c_{\text{abc}})$ where $c_{\text{abc}}$ is a constant representing \code{"abc"} \\
  \multirow{-2}{5.5em}{Strings}   & Any (supertype) & $\texttt{Str}(\texttt{cst})$ where $\texttt{cst}$ is the supertype of constants \\\rowcolor[rgb]{ .906,  .902,  .902}

                                     & Empty list \code{[]} & $\texttt{Lst}(c_{\text{nil}})$ where $c_{\text{nil}}$ is a constant representing \code{[]} \\\rowcolor[rgb]{ .906,  .902,  .902}
                                     & Any (supertype) & $\mu X.\;\texttt{Lst}(\pair\Any X)\vee \texttt{Lst}(c_{\text{nil}})$ \\\rowcolor[rgb]{ .906,  .902,  .902}
                                     & $t$ followed by $l$ & $\texttt{Lst}(\pair{t}{l})$ \\\rowcolor[rgb]{ .906,  .902,  .902}
                                     & \code{$t$*} followed by $l$ & $\mu X.\;\texttt{Lst}(\pair t X)\vee l$ \\\rowcolor[rgb]{ .906,  .902,  .902}
\multirow{-4}{5.5em}{Lists (regexp)} & \code{[ $t_1$ ; $t_2$ ; $t_3$* ]} &  $\texttt{Lst}(\pair{t_1}{\texttt{Lst}(\pair{t_2}{(\mu X.\; \texttt{Lst}(\pair{t_3}{X})\vee \texttt{Lst}(c_{\text{nil}}))})})$ \\
  \bottomrule
\end{NiceTabular}
\renewcommand{\arraystretch}{1.0}
}

\end{table}

As an example, Table~\ref{tab:extensions} proposes some extensions and the corresponding encodings.
The list encoding allows us to support regular expression types purely as an extension. Although they
are isomorphic to recursive products, we can extract from their DNF an automaton representation
of the regular language they capture, which can then be decompiled into a regular expression using
Brzozowski Algebraic Method~\cite{algregexp} before being pretty-printed to the user.
Other extensions have also been implemented through encodings, such as Booleans, floating-point numeric types, and
opaque data types supporting multiple parameters of different variances.

\paragraph{Gradual Types} An important contribution of \cite{phdlanvin} is that
gradual types (that may contain a dynamic component ``\textsf{?}'') can be fully
represented by a pair of static types (its lower and upper materializations). A
type system implementor can simply represent types as a pair of SSTT types without
any modification. This is the approach taken by
\href{https://github.com/elixir-lang/elixir/tree/main/lib/elixir/lib/module/types}{Elixir}
\cite{elixir}.

\section{Evaluation}\label{sec:eval} The SSTT library is implemented in OCaml.
It amounts to 6900 loc, of which 3700 are the core type algebra and subtyping
(Sections~\ref{sec:bdd},~\ref{sec:nodes} and \ref{sec:subtyping}). The code for
the tallying is 470 loc, the rest being extensions
(Section~\ref{sec:extensions}) and the pretty-printer (Appendix~\ref{app:pp}).
The goal of this evaluation is to showcase the benefits of the optimizations we
presented. Specifically, we demonstrate that
\newcommand{\IDESCR}[1]{\normalfont\textit{#1}}
\begin{description}
  \item[\IDESCR{Type representation}:] representing types as BDT with
  semantic simplifications is the right choice;
  \item[\IDESCR{Subtyping}:] our optimized subtyping
  (Alg.~\ref{code:subtyping3}) is a clear improvement w.r.t. the naive one
  (Alg.~\ref{code:subtyping1});
  \item[\IDESCR{Tallying}:] our implementation of tallying
  (Sec.~\ref{sec:tallying}) is a clear improvement w.r.t.
  \cite{polyduce2}.
\end{description}
We also compare SSTT to (to the best of our knowledge) the only known complete
and efficient implementation of set-theoretic types, that is the one found in
the CDuce compiler. We discuss the specific differences between CDuce and our
implementation in the related work section. All experiments in this section can
be reproduced with the companion artifact \cite{stt-artifact}.

\subsection{Experiments}
We used a Linux workstation running Ubuntu 25.04, an Intel
i5-4690@3.50GHz CPU and 32 GB of ram. All the code (whether using SSTT or CDuce) is
compiled in native mode, using OCaml 5.4.1. As we want
to benchmark only the primitive operations offered by SSTT and \emph{not} a
particular type system implementation, we devised the following experiment. In a
first phase we type-checked several program files with the MLSem prototype
\cite{mlsem} that uses SSTT as a type library, and which is an improved version
of \cite{popl24a} (which itself used CDuce as a backend for types). The version
we ran was instrumented to record in JSON files all the tallying instances that
were solved during type-checking. This was done for three corpuses (some
examples are given in Appendix~\ref{app:benchmarks}) with the following
features:
\begin{description}
  \item[\IDESCR{A. Hindley-Milner}] (142 loc, 29 functions, $852$ tallying
  instances) Generic functions in a Hindley-Milner style inspired from the List
  and Map modules of OCaml's standard library. Most functions are recursive and
  involve pattern-matching over list or balanced binary tree types, and their
  type is inferred. The tallying instances generated involve multiple type
  variables (due to parametric polymorphism and recursion), but few unions or
  intersections (unions are only used to encode ADTs with disjoint
  constructors).
  \item[\IDESCR{B. Overloaded}] (221 loc, 69 functions, $5520$ tallying
  instances) Non-recursive and non-generic overloaded functions. For each, a
  type capturing its overloaded behavior is inferred using type narrowing
  techniques. The most overloaded function is an intersection of 26 arrows. The
  tallying instances involve many intersections and unions, but fewer
  type variables.
  \item[\IDESCR{C. HM+Overloaded}] (346 loc, 81 functions, $2335$ tallying
  instances) Recursive, generic and overloaded functions. This corpus involves
  parametric polymorphism, ad-hoc polymorphism, and encodings to model mutable
  data-structures (references, arrays, etc.). The tallying instances generated
  involve unions, intersections, and multiple type variables. This corpus also contains
  about 800 tallying instances that do not have solutions (either because it contains
  purposely ill-typed functions, or because some patterns or type cases become unreachable for
  some domains of overloaded functions).
\end{description}

\paragraph{First Phase, Tallying Instance Generation}
We tested four different flavors of type representation:
\begin{description}
  \item[\IDESCR{BDT}:] types consist of plain reduced BDTs
  \item[\IDESCR{SS}:] types are reduced BDT with semantic simplifications (Sec. \ref{sec:bdd-simpl})
  \item[\IDESCR{HC}:] types are reduced and hash-consed BDT (syntactically equal
  subtrees are shared in memory and set-theoretic operations and subtyping cache
  their results in memoization tables, as is usual for BDD operations)
  \item[\IDESCR{SS+HC}:] types are semantically simplified and hash-consed BDT
\end{description}
For all four configurations, we tried to type-check the three corpuses while
recording the tallying problems solved during type-checking. The result of this
first phase is that, unless semantic simplification is systematic
(modes \textit{SS} and \textit{SS+HC}), inferred types (in particular arrows) get too
large during type-checking, to a point where some tallying instances do not
terminate after a minute. This phenomenon occurs for all three corpuses.
Implementing hash-consing for all the components and layers of the node
structure does not help, though it allows sharing about 50\% of the nodes, thus
reducing the size of some BDTs by making more atoms syntactically equivalent.

Consequently, our benchmark JSON files used in the second phase have been
generated by running the type-checker in a setting where types are always
semantically simplified. The practical impact is that the types stored in JSON
files contain very few redundant atoms (serializing types into a
syntactic form may re-introduce some redundancy).

\paragraph{Second Phase, Tallying Instance Solving}
The result of replaying the JSON files in various modes is given in
Table~\ref{fig:eval}. Each tallying instance in the JSON file is first parsed
into a simple AST (not timed). The ``Building'' line represents the cumulative
time (in s) necessary to build the type representations from the AST for all
instances in the file. The ``Solving'' line reports the cumulative time to
perform tallying and apply the resulting substitution to the initial type. The
``Total'' line is the sum of the two previous times. The ``Slowdown'' line
reports the slowdown of the ``Total'' time w.r.t. the first BDT column
(e.g. for corpus A, type representation SS is 1.61 times slower than BDT).
The ``\# Sol.'' line gives the total
number of tallying solutions found for the corpus. For a given corpus,
we aim to find the smallest number of solutions, as a higher number means that some solutions are redundant.
The ``Size'' line reports the cumulative size of types in the tallying instances and their results. More
specifically, it is the sum of the size of all tallying problems, all
substitution solutions and all result types (applications of the latter to the
former) of the tested file, in bytes. The size of a type is obtained by serializing
it with OCaml's \texttt{Marshal} library. The key point is that sharing is
preserved and therefore, the serialization gives a good approximation of the
\emph{unique} subtrees that constitute a type. The ``Avg. Size'' line is just the ``Size'' line averaged by the number
of entries in the file. The ``Peak Size'' line reports the size in bytes of the
largest type built during the process (including intermediary types built during
tallying and discarded). It is unavailable for CDuce as it is not possible to
obtain it without modifying the internals of the library. Lastly, the
``Timeout'' line reports the number of instances that could not be solved in
under ten seconds. In such cases the reported timings only include the time
for instances that did not timeout.
\begin{table}[h]
  \centering
  \caption{Time performance (in s), number of solutions and size of types (in bytes) for all
  configurations\label{fig:eval}}
  \vspace*{-2mm}
  \scalebox{0.85}{
  \newcommand{\GRAYLINE}{\rowcolor[rgb]{ .906,  .902,  .902}}
  \newcommand{\WHITELINE}{\rowcolor[rgb]{ .906,  .902,  .902}}
  \newcommand{\CMIN}[1]{\cellcolor{blue!20}{#1}}
  \newcommand{\CMAX}[1]{\cellcolor{red!20}{#1}}
\newcommand{\SD}[1]{$\times$\,#1}
  \

\renewcommand{\arraystretch}{1.0} 
\begin{NiceTabular}{ |l|l||c:ccc||c:c||@{}c@{}||c| }[first-row]
  & & \multicolumn{4}{c}{Type representation} &
  \multicolumn{2}{>{\centering\arraybackslash}b{2cm}}{Naive Subtyping} &
  \multicolumn{1}{>{\centering\arraybackslash}b{1cm}}{\,\,\,Naive Tallying}  &
  \\
    \toprule
  Corpus & Measure & \begin{tabular}{@{}c@{}} BDT \\ (ideal) \end{tabular}  & SS$^\star$ & HC & {SS+HC} & BDT & SS & SS &CDuce \\
  \midrule
 \GRAYLINE     &  Building   &  0.021 &  0.065 &  0.051 &  0.203 &  0.023 &  0.078 &  0.071 &  0.035\\
 \GRAYLINE     &  Solving    &  0.210 &  0.307 &  0.392 &  0.760 &  0.321 &  0.486 &  8.722 &  0.446\\
 \GRAYLINE A.    &  Total      &  0.231 &  0.373 &  0.442 &  0.962 &  0.344 &  0.564 &  8.794 &  0.481 \\
 \GRAYLINE      & Slowdown  &      -     & \SD{1.61}   &  \SD{1.91}  &  \SD{4.16} &   \SD{1.49} & \SD{2.44}  & \SD{37.07} & \SD{2.08} \\
 \GRAYLINE     &  \# Sol. &  870 &  870 &  870 &  870 &  870 &  870 &  5244 &  874\\
 \GRAYLINE  852  &  Size       &  3.2M &  3.0M &  2.3M &  2.1M &  3.2M &  3.0M &  11.0M &  3.2M\\
 \GRAYLINE instances    &  Avg. Size  &  3.8k &  3.5k &  2.7k &  2.5k &  3.8k &  3.5k &  12.9k &  3.8k\\
 \GRAYLINE     &  Peak Size  &  18.9k &  9.3k &  11.1k &  6.7k &  18.9k &  9.3k &  9.3k &  N/A\\
 \GRAYLINE     &  Timeout    &  0 &  0 &  0 &  0 &  0 &  1 &  2 &  0\\
               &  Building   &  0.016 &  0.060 &  0.050 &  0.196 &  0.013 &  0.069 &  0.060 &  0.044\\
               &  Solving    &  0.144 &  0.221 &  0.373 &  0.518 &  0.143 &  0.254 &  0.762 &  0.514\\
          B.   &  Total &  0.160 &  0.281 &  0.423 &  0.714 &  0.156 &  0.323 &  0.822 &  0.557\\
               & Slowdown  &      -     & \SD{1.76}   &  \SD{2.64}  &  \SD{3.46} &   \SD{0.97} & \SD{2.02}  & \SD{5.14} & \SD{3.84} \\

          &  \# Sol. &  5790 &  5790 &  5790 &  5790 &  5790 &  5790 &  10676 &  5790\\
         5520      &  Size       &  3.9M &  3.7M &  3.4M &  3.1M &  3.9M &  3.7M &  4.8M &  2.8M\\
         instances  &  Avg. Size  &  715 &  666 &  615 &  563 &  715 &  666 &  869 &  502\\
               &  Peak Size  &  8.3k &  8.5k &  3.8k &  3.8k &  8.3k &  8.5k &  8.5k &  N/A\\
               &  Timeout    &  0 &  0 &  0 &  0 &  0 &  0 &  5 &  0\\
 \GRAYLINE     &  Building   &  0.020 &  0.064 &  0.061 &  0.187 &  0.016 &  0.065 &  0.063 &  0.034\\
 \GRAYLINE    &  Solving    &  0.206 &  0.289 &  0.378 &  0.639 &  0.222 &  0.483 &  15.498 &  0.558\\
 \GRAYLINE C.  &  Total &  0.227 &  0.353 &  0.439 &  0.826 &  0.238 &  0.548 &  15.561 &  0.592\\
 \GRAYLINE      & Slowdown  &      -     & \SD{1.56}   &  \SD{1.93}  &  \SD{3.64} & \SD{1.05}& \SD{2.41}  & \SD{68.5} & \SD{2.61} \\

 \GRAYLINE    &  \# Sol. &  1569 &  1569 &  1569 &  1569 &  1569 &  1569 &  16208 &  1629\\
 \GRAYLINE 2335  &  Size       &  2.6M &  2.3M &  2.2M &  1.9M &  2.6M &  2.3M &  19.2M &  3.5M\\
 \GRAYLINE instances   &  Avg. Size  &  1.1k &  989 &  922 &  806 &  1.1k &  990 &  8.2k &  1.5k\\
 \GRAYLINE     &  Peak Size  &  20.7k &  6.9k &  13.1k &  4.8k &  20.7k &  6.9k &  6.9k &  N/A\\
 \GRAYLINE     &  Timeout    &  0 &  0 &  0 &  0 &  0 &  0 &  5 &  0\\
 \bottomrule
\end{NiceTabular}
\renewcommand{\arraystretch}{1.0}
}

\vspace*{1mm}
\noindent {\small $\star$ : Default mode of the SSTT library}

\end{table}

The table is composed of four parts. The first part, ``Type representation''
fixes the subtyping and tallying to their optimized version and changes the
underlying type representation (for the same four variants described in the
first phase). The second part, ``Subtyping'' uses the textbook subtyping
algorithm (Alg.~\ref{code:subtyping1}) for the two type representation of
reduced BDT and semantically simplified (SS) BDT. The third part, ``Tallying''
disables the constraint simplifications of Section~\ref{sec:tallying}, thus
generating the same number of constraints as in \cite{polyduce2}.
Lastly, the ``CDuce'' column is the one where SSTT is replaced by CDuce and uses
CDuce's types, subtyping and tallying. Some type constructors are not supported
by CDuce and have been encoded (cf. Appendix~\ref{app:benchmarks}).

\subsection{Analysis}

\paragraph{On Mandatory Semantic Simplifications} As the results of the tallying
constraint generation indicate, some form of semantic simplification is needed
when dealing with set-theoretic types. Without these simplifications, BDTs
become full of redundant atoms, to a point that traversing them during tallying
becomes too costly. Instrumenting the first phase we found that some
intermediary types peak at 50MB if not simplified. Simple hash-consing does not
help either. Although it allows sharing some nodes, these nodes are still
present in the BDT and are traversed during tallying. This phenomenon is also
present when using CDuce as a type representation. Indeed, the published
\cite{popl24a} implementation of \cite{popl24}, which uses CDuce's type
data-structures, subtyping and tallying also tries to simplify types
semantically. It does so by turning each type into its DNF and removing
semantically redundant atoms. Without this (inefficient) process, types grow too
large and e.g. typechecking the \emph{flatten} function on lists does not
terminate in a reasonable time\footnote{Our artifact includes a reproducer to
verify this claim.}.

\paragraph{Type Representations} In the first group of columns of
Table~\ref{fig:eval}, the ``BDT'' column plays the role of an ``ideal but
unpractical'' representation. Indeed, since we are in a situation where every
input type was the serialization of a semantically simplified type, then the
tallying instances tested contain very few redundant atoms (serialization may
introduce some). The building time then only consists of syntactically
constructing the reduced BDT. Solving constraints (tallying) which involves
calling the subtyping algorithm and reconstructing finitely many new types is
also the fastest. Unsurprisingly, the internal size of types is larger in BDT
mode but the size and average size do not differ much from the other type
representations. One metric which is particularly bad, however, is the ``Peak''
size of types for corpuses A. and C., where the peak size is respectively two
and three times larger than for the SS representation. This can be explained by
the fact that these corpuses feature polymorphic input and output types, where
many type substitutions occur. These may accumulate during type-checking and
create very large types (in particular arrow types) which then make tallying
impractical (e.g. while type-checking function composition or curried function
applications). On the other hand, with semantic simplification, some cost is
paid upfront during type building or when building intermediary types during
tallying (the SS strategy does not assume that the type has been simplified in
phase 1, which explains the higher building time). However, this longer building
time pays off, as it allows performing tallying reasonably efficiently while
preventing types from growing too much. Indeed, the SS strategy has
consistently the smallest overhead in runtime (between 1.56 and 1.76 times) w.r.t. the
BDT strategy.
Regarding hash-consing (with or without
semantic simplification), it seems that the overhead caused by the look-up in
memoization tables outweighs the benefits of the reduced type size. We therefore
conclude that systematic semantic simplification is a good default strategy
which balances type size with runtime efficiency. Finally, we have not observed
the order over variables and atoms used in BDTs to be a significant factor even
though one can create a type for which the BDT representation is
exponential\footnote{The type $\displaystyle\bigwedge_{i\leq n} ((\alpha_i
\wedge \beta_i) \vee (\lnot\alpha_i \wedge \lnot\beta_i))$ where
$\alpha_i\tleq\alpha_{i+1}$, $\beta_i\tleq\beta_{i+1}$ and
$\alpha_i\tleq\beta_{i}$ yields a BDT exponential in $n$.}. We have not observed
such degenerated cases in our tests.

\paragraph{Subtyping} The second group of columns shows the impact of our
improved subtyping algorithm. First for tallying instances that directly
translate to simple subtyping tests (corpus B), switching to the naive
algorithm shows the opposite effect, the naive algorithm seems slightly faster (in BDT
mode). This is because types are so simple that the caching machinery simply
adds overhead. However, the benefit of caching is real and adds up. For instances
where the subtyping procedure is called repetitively either by tallying or by
the semantic simplification of types, reverting to the naive subtyping algorithm
has a dramatic effect (a simplification test in corpus A. even times out).

\paragraph{Tallying} To test our improvement w.r.t. \cite{polyduce2}, we
deactivated the simplification of constraints (we do not apply $\CSSsimp$ when merging constraint sets). Here the results are clear: it is
the only mode where several examples failed due to timeouts (all examples involve
overloaded functions with 10 arrows or more). Also note the sharp increase in
the ``$\#$ Sol.'' and ``Size'' lines. As soon as tallying computes non-trivial substitutions (corpuses A and C)
the number of redundant solutions found by the naive tallying is 6 to 10 times larger than the reduced
set found by our improved algorithm.

\paragraph{CDuce} Last but not least, our experiments show that our default
configuration for SSTT outperforms CDuce's default one. In CDuce, types are
represented by \emph{lazy BDDs} (see Section~\ref{sec:related}) for which type building is slower than
plain BDT but faster than SS. CDuce's subtyping works by building
an inhabitant of the type, is written in CPS style, and avoids backtracking by
traversing the type in BFS rather than DFS. In hindsight, although it may explore
a subtree only when necessary, it performs a lot of memory allocations of
closures which are detrimental in practice. For the three corpuses, we are respectively
20\%, 50\% and 40\% faster than CDuce.
Last, CDuce also generates more
constraints, although not as much as \cite{polyduce2} as it performs an early
and unpublished version of $\CSSsimp$.

\section{Related Work and Future Work}\label{sec:related}

\paragraph{Set-Theoretic Types in the CDuce Language}
The \href{https://gitlab.math.univ-paris-diderot.fr/cduce/cduce}{CDuce language}
\cite{AlainThesis,cduceicfp} was the first to implement monomorphic
higher-order set-theoretic types (the earlier XDuce \cite{xduce} only featured
first-order, top-level functions and no arrow type), later extended with
parametric polymorphism~\cite{polyduce2}. This implementation supports arrows,
pairs, records, integer intervals, characters and XML types. The overall
architecture of descriptors we use to represent set-theoretic types follows the
one described in~\cite{Cas20}, though the latter does not feature type
variables. More generally, CDuce's internal type representation and subtyping
algorithm were introduced with XML aware programming in mind (its internal type
algebra features XML namespaces and XML document types). And although
\cite{polyduce2} introduced parametric polymorphism and tallying for CDuce,
their implementation was never discussed nor evaluated. Our work provides this
missing piece in the literature, providing efficient implementations of all the relevant algorithms
and data-structures, in an open and modular library.
\paragraph{Set-Theoretic Types in Mainstream Languages}
Two examples of set-theoretic type systems for
mainstream languages are \href{https://github.com/etylizer/etylizer}{Etylizer} \cite{erlang} (for Erlang),
and \href{https://github.com/elixir-lang/elixir/tree/main/lib/elixir/lib/module/types}{Elixir} \cite{elixir}.
Both provide an implementation of set-theoretic types
that supports arrows and a generalized form of records for typing maps. 
While they seem to rely on the same ideas that were already present in CDuce
(some form of BDD to represent types for instance), making a direct comparison
is still not possible, either because their implementation is too slow
\cite{erlang} or in the case of Elixir, because it does not yet support
recursive types nor type variables.

\paragraph{Lazy BDDs} \href{https://gitlab.math.univ-paris-diderot.fr/cduce/cduce/-/blob/dev/types/bdd.ml}{CDuce}
and \href{https://github.com/elixir-lang/elixir/pull/14806}{Elixir} use lazy BDDs
for representing DNF. As explained in \cite{Cas20}, lazy BDDs avoid a potential explosion in size
when repeatedly applying unions, and this until an intersection, difference or negation is performed.
When it happens, the accumulated unions of the lazy BDD must be distributed to compute a plain BDD,
the hope being that the intersection or difference still yields a simpler BDD by canceling some lazy unions.
Our semantic simplification (Section~\ref{sec:bdd-simpl}) can be adapted to work on lazy BDDs,
but doing so increases its complexity and voids Proposition~\ref{prop:bsimple_neg}, which allows us to quickly compute
simplified negations. Overall, even if both approaches aim at keeping type representations small,
they have different trade-offs: our systematic simplification tends to increase the building time of BDDs
but simplifies future manipulations. Lazy BDDs have linear time unions but an increased complexity for
some operations (such as negations).

\paragraph{Syntactic and Algebraic Approaches} While there are clear
\emph{theoretical} differences between semantic subtyping and syntax based ones,
the separation lies elsewhere when it comes to implementation. Either the type
system enjoys only \emph{disjoint} union types, as is the case of MLsub
\cite{mlsub}. In such works, subtyping remains polynomial, since types are
equivalent to deterministic automata, and subtyping amounts to inclusion of
their language, which is polynomial. Or some form of non-disjoint union occurs,
as is the case in the early work of Aiken \cite{aiken91} or more recently
MLStruct \cite{mlstruct}. In both cases, either an exponential blow-up occurs
when computing type operations (e.g. negation) or when deciding subtyping, by
traversing the potentially exponential DNF of the type as we do. Both works
mention in passing that memoizing intermediate results in a mutable table helps
in practice, but do not give any detail. We believe our techniques could be
applied to these works.

\paragraph{SMT-Based Approaches}
Using SMT solvers to decide subtyping has been attempted before. For instance
\cite{semsubsmt} uses an SMT solver to type-check the $D$-minor language. More
recently, \cite{rtts} uses an SMT solver to type-check TypeScript, a typed
dialect of JavaScript which features polymorphism, higher-order and
intersection. While these works show that subtyping can to some extent be
encoded into an SMT problem, it poses several challenges. First,
\cite{semsubsmt} does not feature higher-order nor parametric polymorphism,
which greatly simplifies the encoding. Second, it is not clear whether such
encodings would rely on undecidable theories (and therefore would yield
an incomplete subtyping). Also, as in \cite{rtts}, unification often depends
on some ad-hoc procedure not directly handled by the SMT solver. In that work,
type-checking a splay tree example program (18
functions, 206 loc and no refinement types) takes about 6 seconds which
seems to indicate that, when refinement types are not needed,
then the tailored algorithms we propose are better suited.

\paragraph{Data-Flow Analysis} The knowledgeable reader will remark that the subtyping
algorithm could be cast in the framework of data-flow analysis. It is,
after all, the computation of a fixed point of some monotonic predicate over a
graph (the type, whose vertices are the nodes). For instance, Kildall's
algorithm \cite{Kildall} could be used to compute the
emptiness predicate. The major difference with our approach is that, to the best
of our knowledge, all variants of Kildall's algorithm require the
full graph. In our case, this would entail calling
the $\Psi_\textsf{prod,arrow}$ functions eagerly, generating a potentially
exponential type up-front to test its emptiness.

\paragraph{Conclusion and Future Work} We proposed a modular data-structure to
represent recursive set-theoretic types supporting multiple type constructors,
unions, intersections, negations, and type substitutions. This representation
relies on Binary Decision Trees (BDTs), equipped with a semantic simplification procedure
which eliminates redundancy that accumulates when manipulating types, and we provide
practical improvements to the traditional algorithms for subtyping and tallying.
The performance of SSTT is compared against CDuce, the historical
implementation of set-theoretic types, showing a significant improvement.
While SSTT is already usable, we plan to extend it to support row polymorphism
for records \cite{Cas23}, as well as extensions relevant to dynamic languages
such as Python (such as object types).
\clearpage

\section*{Data Availability Statement}

All the data-structures and algorithms described in this paper have been implemented in the OCaml library SSTT,
available in the software artifact associated with this paper \cite{stt-artifact}.
This artifact contains both the source programs and the generated benchmark files containing the tallying instances,
as well as instructions for replicating the benchmarks.
An extended version of this paper with appendices can be found in the supplementary material.

\begin{acks}
We would like to thank the reviewers and organizers of the OOPSLA'26
conference.  This work was supported by the Czech Ministry of
Education, Youth and Sports under program ERC-CZ, grant agreement
LL2325, as well as the SECUREVAL ANR project n. ANR-22-PECY-0005.
\end{acks}

\bibliographystyle{ACM-Reference-Format}
\bibliography{main}

\clearpage
\appendix

\section{Semantic Subtyping}\label{app:interpretation}

This appendix summarizes the interpretation of set-theoretic types formalized in~\cite{Gesbert2015}
and how it can be used to define semantic subtyping.

\subsection{Type Interpretation}

In order to define subtyping over these types, the idea is to interpret each ground type
(i.e., a type that does not contain type variables) as a set of values of our language.
Then, subtyping can be defined as set containment over the interpretation of types.
Intuitively, each ground type is associated with the set of values having this type:
for instance, the base type $\True$ is interpreted as the singleton containing the constant
$\true$, while the type $\Bool=\True\vee\False$ is interpreted as the set $\{\true,\false\}$.

However, this idea becomes subtler when dealing with arrow types. Although an arrow type
intuitively corresponds to a function (i.e., a $\lambda$-abstraction), interpreting an arrow type
as a set of $\lambda$-abstractions is problematic as it yields a circular reasoning:
determining if a $\lambda$-abstraction is in the interpretation of a type requires
defining a type system, which in turn needs the subtyping relation that we are trying to build.
In order to break this circularity, the
interpretation of types is not defined over values of our language
but over a domain $\Domain$ defined below.
Note that this does not necessarily invalidate the ``types as sets of values'' intuition,
as discussed in \citet[Section~2.7]{gentleintro}.


In addition, we need to define an interpretation for all types, and not only ground ones
(the interpretation domain $\Domain$ should account for type variables).
A simple model was proposed by~\cite{Gesbert2015}.
We succinctly present it in this section.
The reader may refer to~\cite[Section 3.3]{Cas22} for more details. 
\begin{definition}[Interpretation domain~\cite{Gesbert2015}]\label{def:interpretation}
 The \emph{interpretation domain} $ \Domain $ is the set of finite terms $ d $
produced inductively by the following grammar
\begin{align*}
  d & \Coloneqq  c^L \mid (d, d)^L \mid \Set{(d, \domega), \dots, (d, \domega)}^L
    \\
    \domega & \Coloneqq d \mid \Omega
\end{align*}
where $ c $ ranges over the set $ \Constants $ of constants, $L$ ranges over finite sets of type variables,
and where $ \Omega $ is such that $ \Omega \notin \Domain $.
\end{definition}
The elements of $ \Domain $ correspond, intuitively,
to (denotations of) the results of the evaluation of expressions,
labeled by finite sets of type variables. 
In particular, in a higher-order language,
the results of  computations can be functions which, in this model,
are represented by sets of finite relations
of the form $ \Set{(d_1, \domega_1), \dots, (d_n, \domega_n)}^L $,
where $ \Omega $ (which is  not in $ \Domain $)
can appear in second components to signify
that the function fails (i.e., evaluation is stuck) on the
corresponding input. This is implemented by using in the second
projection the meta-variable $\domega$ which ranges over
$ \Domain_\Omega = \Domain \cup \Set{\Omega} $ (we reserve
$d$ to range over $\Domain$, thus excluding $\Omega$).
This constant $\Omega$ is used to ensure that $\Any\to\Any$ is not a
supertype of all function types: if we used $d$ instead of $\domega$,
then every well-typed function could be subsumed to  $\Any\to\Any$
and, therefore, every application could be given the type $\Any$,
independently of its argument as long as this argument is typeable (see Section 4.2 of~\cite{FCB08} for details).
The restriction to \emph{finite} relations corresponds to the intuition
that the denotational semantics of a function is given by the set of
its finite approximations, where finiteness is a restriction necessary
(for cardinality reasons) to give the
semantics to higher-order functions. Finally, the sets of type
variables that label the elements of the domain are used to interpret
type variables: we interpret a type variable
$\alpha$ by the set of all elements that are labeled by $\alpha$, that
is the set $\Set{ d\mid \alpha\in\tags(d)}$ (where we define
$\tags(c^L)=\tags((d, d' )^L)=\tags(\Set{(d_1, \domega_1), \dots,
  (d_n, \domega_n)}^L)=L$). 

We now define the interpretation $ \typinter{t} $ of a type $ t $
(this notation should not be mistaken with the interpretation $\Bint\B$ of a BDT as a type,
defined in Section~\ref{sec:bdd}).
The interpretation $ \typinter{t} $ should satisfy the following equalities,
where  $ \Pf $ denotes the restriction of the powerset to finite
subsets, and $\ConstantsInBasicType{}$ denotes the function
that assigns to each base type the set of constants of that type, so
  that  for every constant $c$ we have $
  c\in \ConstantsInBasicType(\basic {c})$
(we use $\basic{c}$ to denote the base type of the constant $c$):
\begin{align*}
  \typinter{\Empty} & = \emptyset&
  \typinter{\alpha}& =\Set{ d\mid \alpha\in\tags(d)} &
  \typinter{t_1 \lor t_2} & = \typinter{t_1} \cup \typinter{t_2}
  \\
  \typinter{\lnot t} & = \Domain \setminus \typinter{t} &
  \typinter{b} & = \Set{c^L\mid c\in\ConstantsInBasicType(b)} &
  \typinter{\pair {t_1} {t_2}} & = \Set{(d_1,d_2)^L \mid d_1\in\typinter{t_1}, d_2\in\typinter{t_2}} \\
  \typinter{t_1 {\to} t_2} & =
    \{ R^L \mid R \in \Pf(\Domain \times \Domain_\Omega)\text{ s.t. }\forall (d, \domega) \in R. \:d \in \typinter{t_1} \implies \domega \in \typinter{t_2}\}
   \span\span
   \span\span
\end{align*}
Note that, even though we included $\Any$ and the intersection $\wedge$ in the syntax of our types (Definition~\ref{def:types}),
those two can be defined from the other constructors: $\Any=\neg\Empty$ and $t_1\wedge t_2=\neg(\neg t_1 \vee \neg t_2)$
(De Morgan's law). It is easy to see that, with these definitions, we have $\typinter\Any = \Domain$ and $\typinter{t_1\land t_2} = \typinter{t_1}\cap\typinter{t_2}$.
Thus, it is not necessary to define an interpretation for them.

We cannot take the equations above
directly as an inductive definition of $ \typinter{\cdot} $
because types are not defined inductively but coinductively.
Notice however that the contractivity condition of
Definition~\ref{def:types} ensures that the binary relation $\vartriangleright
\,\subseteq\!\Types{\times}\Types$ defined by $t_1 \lor t_2 \vartriangleright
t_i$, $t_1 \land t_2 \vartriangleright
t_i$, $\neg t \vartriangleright t$ is Noetherian. 
This gives an induction principle\footnote{In a nutshell, we can do
proofs and give definitions by induction on the structure of unions and negations---and, thus, intersections---but
arrows, products, and base types are the base cases for the induction.} on $\Types$ that we
use combined with  structural induction on $\Domain$ to give the following definition,
which validates the equalities above.
\begin{definition}[Set-theoretic interpretation of types]\label{def:typinter}
We define a binary predicate $ (d : t) $
(``the element $ d $ belongs to the type $t$''),
where $ d \in \Domain $ and $ t \in \Types $,
by induction on the pair $ (d, t) $ ordered lexicographically.
The predicate is defined as follows:
\begin{align*}
  (c^L : b)  &= c \in \ConstantsInBasicType(b) \\
  (d:\alpha) &= \alpha\in\tags(d) \\
  ((d_1, d_2)^L : \pair {t_1} {t_2} )  &=
    (d_1 : t_1) \mathrel{\mathsf{and}} (d_2 : t_2) \\
  (\Set{(d_1, \domega_1),..., (d_n, \domega_n)}^L : t_1 \to t_2)  &=
   \forall i \in [1.. n] . \:
    \mathsf{if} \: (d_i : t_1) \mathrel{\mathsf{then}} (\domega_i : t_2) \\
  (d : t_1 \lor t_2)  &= (d : t_1) \mathrel{\mathsf{or}} (d : t_2) \\
  (d : \lnot t)  &= \mathsf{not} \: (d : t) \\
  (\domega : t)  &= \mathsf{false} & \text{ otherwise}
  \end{align*}
We define the \emph{set-theoretic interpretation}
$ \typinter{.} : \Types \to \Pd(\Domain) $
as $ \typinter{t} = \{d \in \Domain \mid (d : t)\} $.
\end{definition}

\subsection{Semantic Subtyping}

Now that we have a set-theoretic interpretation of types,
we can define the subtyping preorder and its associated equivalence relation
as follows.

\begin{definition}[Subtyping relation]\label{def:subtyping}
  We define the \emph{subtyping} relation $ \leq $
  and the \emph{subtyping equivalence} relation $ \simeq $
  as
  \(
    t_1 \leq t_2 \iffdef \typinter{t_1} \subseteq \typinter{t_2}\) and   
  \(t_1 \simeq t_2 \iffdef (t_1 \leq t_2) \mathrel{\mathsf{and}} (t_2 \leq t_1)
    \: .
  \)
\end{definition}

This subtyping relation is decidable and is sometimes referred to as \emph{semantic subtyping},
as it is not defined on the syntax of the type but on its interpretation.

For any two type substitutions $\s_1$ and $\s_2$,
we write $\s_1\simeq\s_2$ the pointwise subtyping equivalence of $\s_1$ and $\s_2$.
An important property of the interpretation above is that subtyping is preserved by type substitutions:
\[\forall t_1,t_2,\s.\ t_1\leq t_2\Rightarrow t_1\s\leq t_2\s\]
\clearpage
\section{Pretty Printing of Types}\label{app:pp}

Section~\ref{sec:nodes} defines an internal representation for types
that captures their semantic properties. This semantic representation makes it possible
to implement set-theoretic connectives ($\land$, $\vee$, $\neg$) efficiently,
going from an algebraic representation of a type into a more semantic representation.
However, when printing a set-theoretic type,
we need to go the other direction: going from our
internal representation back into an algebraic one.
We propose in this section a quick description of the challenges
and techniques we implemented to convert types into an algebraic form.

\subsection{Printing of Descriptors}

For each component of a descriptor (arrow, product, constant), we can easily
extract a DNF of atoms directly from their BDT representation.
However, finding a concise algebraic representation for unions of components requires some more work.
The naive approach consisting in systematically printing the union of all components may generate
overly complicated notations: for instance, the type $\Any$ is printed as \code{AnyArrow | AnyPair | AnyConst},
and the type $t\eqdef \neg(\pair\Int\Int)$ is printed as \code{AnyArrow | (AnyPair $\setminus$ (int,int)) | AnyConst}.
Instead, we propose the following algorithm for printing a descriptor $\Descr$ in a more concise way:
\begin{enumerate}
  \item Each component of $\Descr$ is converted into an algebraic representation,
  and these algebraic representations are regrouped inside a union. We simplify this union by eliminating empty clauses.
  This yields an algebraic representation $P_1$ for $\Descr$.
  For instance, for the descriptor of $t$, this yields $P_1\eqdef$\code{AnyArrow | (AnyPair $\setminus$ (int,int)) | AnyConst}.
  \item Independently, the complement of each component of $\Descr$ is converted
  into an algebraic representation, and these algebraic representations are regrouped inside a union.
  For instance, for the descriptor of $t$, this yields \code{EmptyArrow | (int,int) | EmptyConst}.
  Again, we simplify this union by eliminating empty clauses, yielding \code{(int,int)}.
  We obtain an algebraic representation $N$ for the negation of $\Descr$.
  \item We negate the algebraic representation $N$ and perform trivial simplifications
  (such as elimination of double negations), yielding a representation $P_2$ for $\Descr$.
  In our example, we obtain $P_2\eqdef$\code{$\neg$(int,int)}.
  \item We compare the length of $P_1$ and $P_2$ (using any relevant metric, for instance
  the number of nodes in their AST), and return the smaller one (in our case, $P_2$).
\end{enumerate}

\subsection{Printing of (Recursive) Nodes}

Our types are represented by graphs of nodes that may contain cycles (in the case of recursive types).
Thus, when printing a descriptor, it is not enough to recursively call the printer
on each node referenced by its atoms, as this may result in an infinite inlining of recursive types.
Instead, we adopt the following strategy for printing a node $\Node$:
\begin{enumerate}
  \item For each node in the connected component $\deps\Node$ of $\Node$, we compute an algebraic form
  for the top-level structure of its definition (referenced nodes are kept as symbolic references).
  For instance, when printing the type $\mu X.\;(\pair{(\alpha\land(\Int\to\Bool))}{X})\vee\Nil$
  represented in Figure~\ref{fig:nodes}, this leaves us with:\\
  {\centering\code{N1 where N1=(N2,N1)|nil and N2='a\&(N3->N4) and N3=int and N4=bool}}
  \item Then, we try to inline the definition of each node (\code{N1,N2,N3,N4} in our example),
  but only when doing so reduces the length of the whole algebraic representation. This way, recursive
  references will not be inlined. In our example, the nodes \code{N2, N3, N4} are inlined,
  yielding the algebraic expression \code{N1 where N1=('a\&(int->bool),N1)|nil}.
\end{enumerate}\clearpage
\section{Benchmarks}\label{app:benchmarks}

The corpuses have been type-checked using~\cite{mlsem}, a type system based on
the techniques of inference and type narrowing introduced in~\cite{popl24}.

\paragraph{Compatibility with CDuce} The subtyping constraints generated by the type-checker
involve some type constructors that are not supported by CDuce. Thus, when running
the benchmarks with the CDuce backend, the following encodings are used:
\begin{description}
  \item[Tuples] CDuce does not support tuples of arbitrary arity (it only supports pairs and records).
  Thus, a tuple $(t_1,...,t_n)$ is encoded as a closed record $\{ \texttt{card}:n \;;\; \texttt{\_1}:t_1 \;;\; ... \;;\; \texttt{\_n}:t_n \}$.
  \item[Tagged types] CDuce does not support \emph{tagged types}~\cite{mlsem}
  (in our benchmarks, tagged types are used to define constructors of algebraic data types).
  Thus, a tagged type $\texttt{T}(t)$ is encoded as a pair $(\texttt{\_T}, t)$ where
  \texttt{\_T} is a fresh constant for the constructor \texttt{T}.
  \item[Opaque invariant constructors] Our benchmark HM+Overloaded relies on
  \emph{opaque type constructors}~\cite{mlsem} to encode references and mutable arrays.
  In CDuce, we encode a type $\texttt{ref}(t)$ as a pair
  $(\texttt{\_ref}, (t \to \texttt{unit}, \texttt{unit}\to t))$ where \texttt{\_ref} is a fresh constant.
\end{description}

We now include some excerpts from the corpuses.

\subsection{Hindley-Milner Corpus}

This corpus has been type-checked in a configuration where the type inference algorithm does not try
to infer overloaded types for functions, but single arrow types (as in HM systems).
Here is an excerpt (the type inferred for each function is added as a comment):
\begin{stmlcode}[numbers=left,firstnumber=1,xleftmargin=4mm,framexleftmargin=3mm,numbersep=1mm]
let fixpoint = fun f ->
  let delta = fun x ->
     f ( fun  v -> ( x x v ))
   in delta delta
(* (('a -> 'b) -> x1) -> x1 where x1 = 'c & ('a -> 'b) *)

let length_stub length lst =
  if lst is [] then 0 else (length (tl lst))+1
(* ('a & [ any* ] -> int) -> [] | any::('a & [ any* ]) -> int *)

let length = fixpoint length_stub
(* [ any* ] -> int *)
\end{stmlcode}

\noindent Example of tallying instance generated when type-checking this excerpt:
\begin{verbatim}
"vars": [ "'a", "'b", "'c", "'d", "'e" ],
"mono": [],
"constr": [
    [
    "(('a -> 'b) -> 'c & ('a -> 'b)) -> 'c & ('a -> 'b)",
    "(('d & lst(x1) -> int) -> lst(tuple0 | (any, 'd & lst(x1))) -> int) -> 'e
        where x1 = tuple0 | (any, lst(x1))"
    ]
]
\end{verbatim}

\subsection{Overloaded Corpus}

This corpus has been type-checked in a configuration where the type inference algorithm tries
to infer overloaded types for functions (except when the domain of the function is explicitly given).
Here is an excerpt (the type inferred for each function is added as a comment):
\begin{stmlcode}[numbers=left,firstnumber=1,xleftmargin=4mm,framexleftmargin=3mm,numbersep=1mm]
let succ x =
  match x with
  | 0 -> 1 | 1 -> 2 | 2 -> 3 | 3 -> 4 | 4 -> 5
  | 5 -> 6 | 6 -> 7 | 7 -> 8 | 8 -> 9 | 9 -> 10
  | 10 -> 11 | 11 -> 12 | 12 -> 13 | 13 -> 14 | 14 -> 15
  | 15 -> 16 | 16 -> 17 | 17 -> 18 | 18 -> 19 | 19 -> 20
  | 20 -> 21 | 21 -> 22 | 22 -> 23 | 23 -> 24 | 24 -> 25
  | _ -> 0
  end
(* (0 -> 1) & (1 -> 2) & ... & (24 -> 25) & (~(0..24) -> 0) *)
let test (y:(5..15)) = succ (succ (succ (succ (succ y))))
(* (5..15) -> (10..20) *)
\end{stmlcode}

\noindent Example of tallying instance generated when type-checking this excerpt:
\begin{verbatim}
"vars": [ "'a" ],
"mono": [],
"constr": [
    [
    "(1 -> 2) & (2 -> 3) & (3 -> 4) & (4 -> 5)
    & (5 -> 6) & (6 -> 7) & (7 -> 8) & (8 -> 9) & (9 -> 10)
    & (10 -> 11) & (11 -> 12) & (12 -> 13) & (13 -> 14) & (14 -> 15)
    & (15 -> 16) & (16 -> 17) & (17 -> 18) & (18 -> 19) & (19 -> 20)
    & (20 -> 21) & (21 -> 22) & (22 -> 23) & (23 -> 24) & (24 -> 25)
    & (~(0..24) -> 0) & (0 -> 1)",
    "(5..15) -> 'a"
    ]
]
\end{verbatim}

\subsection{HM+Overloaded Corpus}

This corpus has been type-checked in a configuration where the type inference algorithm
performs type narrowing, and where mutable data structures (arrays, references) are encoded as opaque data types.
Here is an excerpt (the type inferred for each function is added as a comment):
\begin{stmlcode}[numbers=left,firstnumber=1,xleftmargin=4mm,framexleftmargin=3mm,numbersep=1mm]
let filter_imp (f:('a -> bool) & ('b -> false)) (arr:array('a|'b)) =
  let res = array () in
  let mut i = 0 in
  while i < (len arr) do
    let e = arr[i] in
    if f e do push res e end ;
    i := i + 1
  end ;
  return res
(* ('b -> false)&('a -> bool) -> array('b|'a) -> array('a\'b | 'c) *)
\end{stmlcode}

\noindent Example of tallying instance generated when type-checking this excerpt:
\begin{verbatim}
"vars": [ "'a", "'b", "'c" ],
"mono": [ "'a", "'b" ],
"constr": [
    [
    "('b -> bool(false)) & ('a -> bool(true | false))", "'a \\ 'b -> 'c"
    ]
]
\end{verbatim}
\clearpage
\section{Proofs for Section~\ref{sec:bdd-simpl}}\label{app:proofs}


We recall the simplification definition for reference.
Given a BDT $\B = \BN\atom\Bp\Bm$ and a type context $t$,
let ${\Bp}' = \Bsimpl{t\land\atom}{\Bp}$, ${\Bm}' = \Bsimpl{t\land\neg\atom}{\Bm}$,
and $\B' = \BN\atom{{\Bp}'}{{\Bm}'}$.  Then:
\[
  \Bsimpl{t}{\BL{\leaf}} = \BL{\leaf}
  \qquad\text{and}\qquad
  \Bsimpl{t}{\BN\atom\Bp\Bm} =
  \begin{cases}
    {\Bp}' & \text{if } t\land\Bint{{\Bp}'}\simeq t\land\Bint{\B'}\\
    {\Bm}' & \text{else if } t\land\Bint{{\Bm}'}\simeq t\land\Bint{\B'}\\
    \B'   & \text{otherwise.}
  \end{cases}
\]
All three proofs establish a stronger invariant for $\Bsimpl{t}{\B}$ with
an arbitrary context $t$, from which the corresponding proposition follows by
taking $t = \Any$.

\subsection*{Proposition~\ref{prop:correctness} (Correctness)}

\begin{proof}
We prove by structural induction on $\B$ the following invariant:
\begin{equation}\label{inv:correctness}
  \text{for any type } t,\qquad
  t \land \Bint{\Bsimpl{t}{\B}} \;\simeq\; t \land \Bint{\B}.
\end{equation}

\noindent\textit{Leaf case.}\;
$\B = \BL{\leaf}$.  Then $\Bsimpl{t}{\BL{\leaf}} = \BL{\leaf}$, so the claim
is immediate.

\smallskip
\noindent\textit{Node case.}\;
$\B = \BN\atom\Bp\Bm$.  Set ${\Bp}' = \Bsimpl{t\land\atom}{\Bp}$,
${\Bm}' = \Bsimpl{t\land\neg\atom}{\Bm}$, and $\B' = \BN\atom{{\Bp}'}{{\Bm}'}$.
The induction hypotheses give
\begin{align}
  (t\land\atom)\land\Bint{{\Bp}'} &\;\simeq\; (t\land\atom)\land\Bint{\Bp}, \label{eq:ih+}\\
  (t\land\neg\atom)\land\Bint{{\Bm}'} &\;\simeq\; (t\land\neg\atom)\land\Bint{\Bm}. \label{eq:ih-}
\end{align}
Expanding $\Bint{\B'}$ by definition and distributing $t$ over the disjunction:
\begin{align*}
  t\land\Bint{\B'}
    &= t\land\bigl[(\Bint\atom\land\Bint{{\Bp}'})\vee(\neg\Bint\atom\land\Bint{{\Bm}'})\bigr]\\
    &= (t\land\atom\land\Bint{{\Bp}'})\vee(t\land\neg\atom\land\Bint{{\Bm}'})\\
    &\simeq (t\land\atom\land\Bint{\Bp})\vee(t\land\neg\atom\land\Bint{\Bm})
      \qquad\text{by \eqref{eq:ih+} and \eqref{eq:ih-}}\\
    &= t\land\Bint{\B}.
\end{align*}
Each branch of $\Bsimpl{t}{\B}$ therefore satisfies the invariant:
\begin{itemize}
  \item If $t\land\Bint{{\Bp}'}\simeq t\land\Bint{\B'}$:
    $\Bsimpl{t}{\B}={\Bp}'$, and
    $t\land\Bint{{\Bp}'} \simeq t\land\Bint{\B'} \simeq t\land\Bint{\B}$.
  \item If $t\land\Bint{{\Bm}'}\simeq t\land\Bint{\B'}$:
    $\Bsimpl{t}{\B}={\Bm}'$, and
    $t\land\Bint{{\Bm}'} \simeq t\land\Bint{\B'} \simeq t\land\Bint{\B}$.
  \item Otherwise: $\Bsimpl{t}{\B}=\B'$, and $t\land\Bint{\B'} \simeq t\land\Bint{\B}$.
\end{itemize}
Taking $t = \Any$:
$\Bint{\Bsimple{\B}} \simeq \Any\land\Bint{\Bsimple{\B}} \simeq
\Any\land\Bint{\B} \simeq \Bint{\B}$.
\end{proof}

\subsection*{Proposition~\ref{prop:bsimple_neg} (Negation)}

\begin{proof}
We first establish an auxiliary lemma.

\smallskip
\noindent\textbf{Lemma.}\;
\textit{For any BDT $\B$, $\;\Bint{\neg\B} \simeq \neg\Bint{\B}$.}

\noindent\textit{Proof.}\;
By induction on $\B$.
The leaf case is immediate: $\Bint{\neg\BL{\leaf}} = \Bint{\BL{\neg\leaf}} =
\neg\Bint{\leaf}$.
For the node case $\B = \BN\atom\Bp\Bm$, note that $\Bint\atom$ and $\neg\Bint\atom$
partition the universe.  Restricting the desired equality to each part:
\begin{align*}
  \Bint\atom \land \neg\Bint{\B}
    &= \Bint\atom \land \neg\Bint{\Bp}
     \;\simeq\; \Bint\atom \land \Bint{\neg\Bp}
     \qquad\text{(IH on $\Bp$),}\\
  \neg\Bint\atom \land \neg\Bint{\B}
    &= \neg\Bint\atom \land \neg\Bint{\Bm}
     \;\simeq\; \neg\Bint\atom \land \Bint{\neg\Bm}
     \qquad\text{(IH on $\Bm$).}
\end{align*}
Combining: $\neg\Bint{\B}
= (\Bint\atom\land\Bint{\neg\Bp})\vee(\neg\Bint\atom\land\Bint{\neg\Bm})
= \Bint{\BN\atom{\neg\Bp}{\neg\Bm}}
= \Bint{\neg\B}$.\hfill$\square$

\smallskip
We now prove by structural induction on $\B$ the following invariant:
\begin{equation}\label{inv:neg}
  \text{for any type } t,\qquad
  \neg\Bsimpl{t}{\B} \;=\; \Bsimpl{t}{\neg\B}
  \quad\text{(syntactic equality).}
\end{equation}

\noindent\textit{Leaf case.}\;
$\neg\Bsimpl{t}{\BL{\leaf}} = \neg\BL{\leaf} = \BL{\neg\leaf}
= \Bsimpl{t}{\BL{\neg\leaf}} = \Bsimpl{t}{\neg\BL{\leaf}}$.

\smallskip
\noindent\textit{Node case.}\;
$\B = \BN\atom\Bp\Bm$.
Set ${\Bp}' = \Bsimpl{t\land\atom}{\Bp}$, ${\Bm}' = \Bsimpl{t\land\neg\atom}{\Bm}$,
$\B' = \BN\atom{{\Bp}'}{{\Bm}'}$.
Since $\neg\B = \BN\atom{\neg\Bp}{\neg\Bm}$, the computation of
$\Bsimpl{t}{\neg\B}$ proceeds as follows:
\begin{itemize}
  \item Positive subtree:\; $\Bsimpl{t\land\atom}{\neg\Bp} = \neg\Bsimpl{t\land\atom}{\Bp} = \neg{\Bp}'$\; (IH on $\Bp$).
  \item Negative subtree:\; $\Bsimpl{t\land\neg\atom}{\neg\Bm} = \neg\Bsimpl{t\land\neg\atom}{\Bm} = \neg{\Bm}'$\; (IH on $\Bm$).
  \item Combined node:\; $\BN\atom{\neg{\Bp}'}{\neg{\Bm}'} = \neg\B'$.
\end{itemize}
It remains to show that the branching conditions for $\Bsimpl{t}{\neg\B}$ agree
with those for $\Bsimpl{t}{\B}$.  By the auxiliary lemma,
$\Bint{\neg{\Bp}'} = \neg\Bint{{\Bp}'}$ and $\Bint{\neg\B'} = \neg\Bint{\B'}$,
so the first condition for $\Bsimpl{t}{\neg\B}$ reads
$t\land\neg\Bint{{\Bp}'} \simeq t\land\neg\Bint{\B'}$.
Since complementation within $t$ is order-reversing and involutive,
\[
  t\land X \simeq t\land Y
  \;\Longleftrightarrow\;
  t\land\neg X \simeq t\land\neg Y,
\]
this condition holds if and only if $t\land\Bint{{\Bp}'}\simeq t\land\Bint{\B'}$,
i.e., the first condition for $\Bsimpl{t}{\B}$.  The second condition is
analogous.  The three branches are therefore taken in the same order, and in
each of them $\neg\Bsimpl{t}{\B} = \Bsimpl{t}{\neg\B}$:
\begin{itemize}
  \item First condition holds:\;
    $\Bsimpl{t}{\B} = {\Bp}'$ and $\Bsimpl{t}{\neg\B} = \neg{\Bp}'$.
  \item Second condition holds:\;
    $\Bsimpl{t}{\B} = {\Bm}'$ and $\Bsimpl{t}{\neg\B} = \neg{\Bm}'$.
  \item Otherwise:\;
    $\Bsimpl{t}{\B} = \B'$ and $\Bsimpl{t}{\neg\B} = \neg\B'$.
\end{itemize}
Taking $t = \Any$ yields the proposition.
\end{proof}

\subsection*{Proposition~\ref{prop:bsimple_empty} (Emptiness)}

\begin{proof}
We prove by structural induction on $\B$ the following invariant:
\begin{equation}\label{inv:empty}
  \text{for any type } t \not\simeq \Empty,\qquad
  t\land\Bint{\B} \simeq \Empty
  \;\Longleftrightarrow\;
  \Bsimpl{t}{\B} = \BE.
\end{equation}

\noindent\textit{Leaf case.}\;
$\B = \BL{\leaf}$, so $\Bsimpl{t}{\BL{\leaf}} = \BL{\leaf}$.
We consider the two leaves of the Boolean leaf lattice $\Leaves = \{\LE, \LA\}$:
\begin{itemize}
  \item $\leaf = \LE$:\; $t\land\Bint{\BL{\LE}} = t\land\Empty = \Empty$, and $\BL{\LE} = \BE$.
    Both sides hold.
  \item $\leaf = \LA$:\; $t\land\Bint{\BL{\LA}} = t\land\Any = t \not\simeq \Empty$
    (by hypothesis), and $\BL{\LA} \neq \BE$.
    Both sides fail.
\end{itemize}

\smallskip
\noindent\textit{Node case.}\;
$\B = \BN\atom\Bp\Bm$.
Set ${\Bp}' = \Bsimpl{t\land\atom}{\Bp}$, ${\Bm}' = \Bsimpl{t\land\neg\atom}{\Bm}$,
$\B' = \BN\atom{{\Bp}'}{{\Bm}'}$.

\smallskip
\noindent\textit{Direction $(\Leftarrow)$.}\;
If $\Bsimpl{t}{\B} = \BE$, then by Proposition~\ref{prop:correctness},
\[
  t\land\Bint{\B} \;\simeq\; t\land\Bint{\Bsimpl{t}{\B}}
  = t\land\Bint{\BE} = t\land\Empty = \Empty.
\]

\smallskip
\noindent\textit{Direction $(\Rightarrow)$.}\;
Suppose $t\land\Bint{\B} \simeq \Empty$, i.e.,
\begin{equation}\label{eq:hyp-empty}
  (t\land\atom\land\Bint{\Bp})\vee(t\land\neg\atom\land\Bint{\Bm}) \;\simeq\; \Empty.
\end{equation}
This implies $(t\land\atom)\land\Bint{\Bp}\simeq\Empty$ and
$(t\land\neg\atom)\land\Bint{\Bm}\simeq\Empty$.
Since $t\not\simeq\Empty$ and $t\simeq(t\land\atom)\vee(t\land\neg\atom)$,
at most one of $t\land\atom$ and $t\land\neg\atom$ is empty.
We distinguish three sub-cases.

\begin{itemize}
  \item \textbf{Both $t\land\atom\not\simeq\Empty$ and $t\land\neg\atom\not\simeq\Empty$:}\;
    By the induction hypothesis (each context is non-empty):
    ${\Bp}'= \BE$ and ${\Bm}'= \BE$.
    The reduction property ($\BN\atom{\B}{\B} \leadsto \B$) then gives $\B' = \BE$.
    The first condition $t\land\Bint{{\Bp}'} \simeq t\land\Bint{\B'}$
    reduces to $\Empty \simeq \Empty$, so $\Bsimpl{t}{\B} = {\Bp}' = \BE$.

  \item \textbf{$t\land\atom\simeq\Empty$} (so $t\simeq t\land\neg\atom\not\simeq\Empty$):\;
    From \eqref{eq:hyp-empty}, $(t\land\neg\atom)\land\Bint{\Bm}\simeq\Empty$.
    By the induction hypothesis, ${\Bm}' = \BE$.
    Moreover,
    \[
      t\land\Bint{\B'} = (t\land\atom\land\Bint{{\Bp}'})\vee(t\land\neg\atom\land\Bint{{\Bm}'})
      \simeq \Empty\vee\Empty = \Empty.
    \]
    In particular $t\land\Bint{{\Bm}'} = \Empty \simeq t\land\Bint{\B'}$,
    so the second condition triggers and $\Bsimpl{t}{\B} = {\Bm}' = \BE$.

  \item \textbf{$t\land\neg\atom\simeq\Empty$} (so $t\simeq t\land\atom\not\simeq\Empty$):\;
    Symmetric to the previous case: by the induction hypothesis ${\Bp}' = \BE$, and
    $t\land\Bint{{\Bp}'} = \Empty \simeq t\land\Bint{\B'}$,
    so $\Bsimpl{t}{\B} = {\Bp}' = \BE$.
\end{itemize}
Taking $t = \Any$ (which satisfies $\Any \not\simeq \Empty$) yields the proposition.
\end{proof}
\clearpage
\section{Tallying: Constraint Sets Subsumption Soundness}\label{app:subsumption}

This appendix proves that constraint set subsumption
(Definition~\ref{def:subsumption-improved}) is \emph{sound}:
if $C_1 \sqsubseteq C_2$, then every substitution satisfying $C_1$ also
satisfies $C_2$.  In other words, a more restrictive constraint set can
always be substituted for a less restrictive one without losing any solution.

\subsection*{Definitions}

We first recall that a normalized constraint set $C$ collects upper and lower
bounds on type variables: a triple $(s, \alpha, t) \in C$ requires that any
solution $\sigma$ instantiate $\alpha$ between $s$ and $t$, taking into
account how $\sigma$ instantiates the other variables present in $s$ and $t$.

\begin{definition}[Solution]
  A substitution $\sigma$ \emph{satisfies} a normalized constraint set $C$,
  written $\sigma \models C$, if for every $(s, \alpha, t) \in C$,
  \[
    s\sigma \;\leq\; \alpha\sigma \;\leq\; t\sigma.
  \]
\end{definition}

\subsection*{Key Lemma}

The weakening and strengthening operators $\tub{u}{C}$ and $\tlb{u}{C}$
(Definition~\ref{def:subsumption-improved}) replace the top-level type variables of $u$
by their bounds from $C$ in a monotone (resp.\ anti-monotone) fashion.  The
crucial semantic property is that both operators are transparent under any
solution of $C$.

\begin{lemma}[Semantic transparency]\label{lem:tub-tlb}
  For any type $u$, normalized constraint set $C$, and substitution $\sigma$
  with $\sigma \models C$:
  \[
    \tub{u}{C}\sigma \;=\; u\sigma
    \qquad\text{and}\qquad
    \tlb{u}{C}\sigma \;=\; u\sigma.
  \]
\end{lemma}

\begin{proof}
  We prove both claims simultaneously by induction on $C$.

  \noindent\textit{Base case.}
  $C = \varnothing$.  Then $\tub{u}{\varnothing} = u = \tlb{u}{\varnothing}$
  by definition, so the claims are immediate.

  \noindent\textit{Inductive step.}
  $C = \{(s, \alpha, t)\} \cup C'$ with $\alpha \tleq \alpha'$ for all
  $\alpha' \in \dom{C'}$.  By definition:
  \begin{align*}
    \tub{u}{C} &= \tub{u\tsubsp{\alpha}{\alpha\vee s}\tsubsm{\alpha}{\alpha\land t}}{C'},\\
    \tlb{u}{C} &= \tlb{u\tsubsp{\alpha}{\alpha\land t}\tsubsm{\alpha}{\alpha\vee s}}{C'}.
  \end{align*}

  Since $\sigma \models C$ we have, in particular,
  $s\sigma \leq \alpha\sigma \leq t\sigma$.  For any occurrence of $\alpha$
  in $u$, both substitution pairs evaluate to $\alpha\sigma$ under $\sigma$:
  $(\alpha \vee s)\sigma = \alpha\sigma \vee s\sigma = \alpha\sigma$
  (since $s\sigma \leq \alpha\sigma$) and
  $(\alpha \land t)\sigma = \alpha\sigma \land t\sigma = \alpha\sigma$
  (since $\alpha\sigma \leq t\sigma$).

  Therefore, letting $u_1 = u\tsubsp{\alpha}{\alpha\vee s}\tsubsm{\alpha}{\alpha\land t}$
  and $u_2 = u\tsubsp{\alpha}{\alpha\land t}\tsubsm{\alpha}{\alpha\vee s}$,
  we have $u_1\sigma = u\sigma$ and $u_2\sigma = u\sigma$.

  Since $C' \subseteq C$, we have $\sigma \models C'$.  The induction
  hypothesis for $\tub{\cdot}{\cdot}$ gives $\tub{u_1}{C'}\sigma = u_1\sigma = u\sigma$,
  and the induction hypothesis for $\tlb{\cdot}{\cdot}$ gives
  $\tlb{u_2}{C'}\sigma = u_2\sigma = u\sigma$.
\end{proof}

\subsection*{Main Theorem}

\begin{theorem}[Subsumption soundness]\label{thm:subsumption-soundness}
  Let $C_1$ and $C_2$ be normalized constraint sets with
  $C_1 \sqsubseteq C_2$ (Definition~\ref{def:subsumption-improved}).
  For any substitution $\sigma$ with $\sigma \models C_1$,
  we have $\sigma \models C_2$.
\end{theorem}

\begin{proof}
  Take any $(s, \alpha, t) \in C_2$.  Let $(s', t') = C_1(\alpha)$.
  We must show $s\sigma \leq \alpha\sigma \leq t\sigma$.

  By definition of $C_1(\alpha)$:
  \begin{itemize}
    \item If $\alpha \in \dom{C_1}$: the triple $(s', \alpha, t') \in C_1$, so
      $\sigma \models C_1$ gives $s'\sigma \leq \alpha\sigma \leq t'\sigma$.
    \item If $\alpha \notin \dom{C_1}$: $(s', t') = (\Empty, \Any)$ by
      convention, so $s'\sigma = \Empty \leq \alpha\sigma$ and
      $\alpha\sigma \leq \Any = t'\sigma$ trivially.
  \end{itemize}
  In both cases $s'\sigma \leq \alpha\sigma \leq t'\sigma$.

  The subsumption condition $C_1 \sqsubseteq C_2$ at $(s, \alpha, t)$ gives:
  \[
    \tlb{s}{C_1} \;\leq\; s' \qquad\text{and}\qquad t' \;\leq\; \tub{t}{C_1}.
  \]

  \noindent\textbf{Lower bound.}
  \[
    s\sigma \;=\; \tlb{s}{C_1}\sigma \;\leq\; s'\sigma \;\leq\; \alpha\sigma.
  \]
  The equality is Lemma~\ref{lem:tub-tlb}.  The second inequality holds because
  applying a substitution preserves subtyping (Proposition~\ref{prop:subst},
  i.e.\ $\tlb{s}{C_1} \leq s'$ implies $\tlb{s}{C_1}\sigma \leq s'\sigma$).
  The last inequality is established above.

  \noindent\textbf{Upper bound.}
  \[
    \alpha\sigma \;\leq\; t'\sigma \;\leq\; \tub{t}{C_1}\sigma \;=\; t\sigma.
  \]
  The first inequality is established above.  The second one holds because
  $t' \leq \tub{t}{C_1}$ implies $t'\sigma \leq \tub{t}{C_1}\sigma$
  (Proposition~\ref{prop:subst}).  The equality is Lemma~\ref{lem:tub-tlb}.

  Since $s\sigma \leq \alpha\sigma \leq t\sigma$ for every
  $(s, \alpha, t) \in C_2$, we conclude $\sigma \models C_2$.
\end{proof}

\noindent
The theorem justifies the use of $\sqsubseteq$ to eliminate redundant
constraint sets in $\CSSsimp$: if $C_1 \sqsubseteq C_2$ then every
substitution produced from $C_1$ already satisfies $C_2$, so $C_2$
contributes no new solutions and can safely be discarded.
\clearpage

\end{document}
\endinput